\documentclass[fleqn,usenatbib]{mnras}

\usepackage{newtxtext} 
\usepackage{amsmath}
\usepackage{makecell}
\usepackage{booktabs}
\usepackage{pdflscape}
\usepackage{lineno}
\usepackage{url}  
\usepackage{enumitem}

\usepackage[T1]{fontenc}
\usepackage{CJK}

\DeclareRobustCommand{\VAN}[3]{#2}
\let\VANthebibliography\thebibliography
\def\thebibliography{\DeclareRobustCommand{\VAN}[3]{##3}\VANthebibliography}

\usepackage{graphicx}	
\usepackage{amsmath}	
\usepackage{amssymb}	

\def\kpc{{\rm\,kpc\,}}
\def\Mpc{{\rm\,Mpc\,}}
\def\pc{{\rm\,pc\,}}
\def\msun{{\rm\,M$_{\odot}$\,}}

\defcitealias{qu2023}{Qu23}

\title[ViT-based Local Volume dwarf galaxy Identificationin (VIDA) in the CSST survey]{ViT-based Local Volume dwarf galaxy Identificationin (VIDA) in the CSST survey}

\begin{document}
\begin{CJK*}{UTF8}{gkai}  
\author[Qu \& Yuan et al.]{
Han Qu (曲涵)$^{1,2}$\thanks{E-mail: qvhan@pmo.ac.cn}
Zhen Yuan (袁珍)$^{3}$\thanks{corresponding author, E-mail: zhen.yuan@nju.edu.cn}
Chengliang Wei (韦成亮)$^{1}$, 
Chao Liu (刘超) $^{4,5}$, 
Jiang Chang (常江)$^{1}$, 
\newauthor 
Guoliang Li (李国亮)$^{1}$,
Nicolas F. Martin$^{6, 7}$,
Chaowei Tsai (蔡肇伟)$^{4,8,9}$,
Shi Shao (邵实)$^{4}$,
Yu Luo (罗煜)$^{10}$,
\newauthor
Ran Li (李然)$^{11}$, 
Xi Kang (康熙)$^{12}$, 
Xiangxiang Xue (薛香香)$^{4}$, 
Zhou Fan (范舟)$^{4}$  
\\
$^{1}$Purple Mountain Observatory, Chinese Academy of Sciences, Nanjing 210008, China\\
$^{2}$School of Astronomy and Space Sciences, University of Science and Technology of China, Hefei 230026, China\\
$^{3}$School of Astronomy and Space Science, Nanjing University, Nanjing 210093, China\\
$^{4}$National Astronomical Observatories, Chinese Academy of Sciences, 20A Datun Road, Chaoyang District, Beijing 100012, China\\
$^{5}$Research Center for Astronomical Computing, Zhejiang Laboratory, Hangzhou 311121, China\\
$^{6}$Université de Strasbourg, CNRS, Observatoire Astronomique de Strasbourg, UMR 7550, F-67000 Strasbourg, France\\
$^{7}$Max-Planck-Institut f\"ur Astronomie, K\"onigstuhl 17, D-69117, Heidelberg, Germany \\
$^{8}$Institute for Frontiers in Astronomy and Astrophysics, Beijing Normal University,  Beijing 102206, China\\
$^{9}$School of Astronomy and Space Science, University of Chinese Academy of Sciences, Beijing 100049, China\\
$^{10}$School of Physics and Information Science, Hunan Normal University, Changsha 410081, China\\
$^{11}$Department of Astronomy, Beijing Normal University, Beijing 100875, China\\
$^{12}$Zhejiang University-Purple Mountain Observatory Joint Research Center for Astronomy, Zhejiang University, Hangzhou 310027, China\\
}

\date{Accepted XXX. Received YYY; in original form ZZZ}

\pubyear{2015}

\label{firstpage}
\pagerange{\pageref{firstpage}--\pageref{lastpage}}
\maketitle
\begin{abstract}

Identifying dwarf galaxies within the Local Volume is crucial for constraining the luminosity function of satellite galaxies in the nearby universe. 
We report the detection capabilities of dwarf galaxies within the Local Volume using the Chinese Space Station Telescope (CSST). Based on the simulated imaging data of CSST, we develop a detection and classification pipeline that combines traditional image-based search techniques with advanced machine learning classification models. The simulated Local Volume dwarf galaxies can be identified using a pre-processing method for "extended source detection", followed by classification with a pretrained ViT-Base model. This pipeline achieves a true positive rate (TPR) exceeding 85\% with a false positive rate (FPR) of only 0.1\%.
We quantify the detection completeness of Local Volume dwarf galaxies across a three-dimensional parameter space defined by absolute magnitude ($M_V$), half-light radius ($R_h$), and heliocentric distance, based on simulated single-exposure CSST wide-field imaging survey data.
For unresolved or semi-resolved dwarf galaxies, our method achieves a significantly deeper absolute magnitude detection limit compared to catalog-based approaches, reaching $M_V = -7$ within 10 \Mpc.
By combining this image-based approach with traditional stellar catalog-based "matched filter" techniques, our automated framework established in this work can identify dwarf galaxies within 20 \Mpc for the CSST mission.

\end{abstract} 
\end{CJK*}



\section{Introduction}

The faint end of satellite galaxy luminosity functions are sensitive to cosmological models with different dark matter properties \citep{Governato2015,Forouhar2022,Dekker2022}. The satellite galaxies around the Milky Way (MW) and the Andromeda galaxy (M31) are often studied for their satellite luminosity functions \citep{Koposov2008, Homma2024, Martin2016, am2022pandas,Amandine2025}. Constraining cosmological models on small scales requires a well-surveyed dwarf galaxy population below the current detection limit (M$_{V}$ $>$ -5). It is also important to broaden satellite luminosity function investigations beyond the Local Group to the Local Volume ($\sim$20\Mpc) to encompass a statistically significant sample of galaxy systems. Substantial progress has been made with recent surveys \citep{Carlsten&2020, Bennet2020,Davis2021,Engler2021,Kanehisa2024,Gozman2024}, such as the Local Volume Legacy survey \citep{Lee2008}, the Dragonfly Nearby Galaxies survey \citep{Merritt2016}, the Exploration of Local VolumE Satellites (ELVES) survey \citep{Carlsten2022}, and the SAGA survey \citep{Tollerud2022}. So far, the observed faint end of the satellite luminosity in the Local Volume has been pushed to $M_{V}<-9$ \citep{Crosby2023}.

The fourth-generation of large-scale survey telescopes, such as the Legacy Survey of Space and Time (LSST; \citealt{LSST2019}), the Euclid Space Telescope (Euclid; \citealt{Euclid&2022,Euclid&2025}), and the Chinese Space Station Telescope (CSST; \citealt{zhan2021}) will provide deeper and wider-area imaging survey data. These advance can significantly improve satellite galaxy searches within the Local Volume. The 2-m CSST, set to launch in the near future, features a main survey camera with a wide 1.1 $\rm deg^{2}$ field-of-view and a spatial resolution of $\sim$ 0.15" . It covers near-ultraviolet to near-infrared wavelengths with $NUV$ and $u$, $g$, $r$, $i$, $z$, $y$ filters. The limiting magnitude in the $r$ band reaches 26 mag in CSST's main survey. All these capabilities would make CSST a powerful tool to detect dwarf galaxies in the Local Volume.

Two methods for searching for nearby dwarf galaxies are often used: catalog-based\citep{Koposov2008,Walsh2009} and image-based methods\citep{Carlsten&2020,Davis&2021}. For dwarf galaxies within 10\Mpc, their member stars are spatially resolvable with space telescopes, making catalog-based methods suitable. The overall procedure of this approach is to first select stars along the giant branch on color-magnitude-diagram (CMD). Then, the significance of the overdensity of these selected stars are evaluated, which allows for the identification of dwarf galaxy candidates. Classic methods such as the "matched filter" technique \citep{Drlica-Wagner2015, Laevens2015a, Simon2019} and those based on likelihood estimators \citep{Martin2013} have led to successful discoveries of Local Group dwarf galaxies. In our previous work, we utilize the classic approach and evaluated the detection limits of dwarf galaxies with the CSST \citep{qu2023}. For more distant dwarf galaxies in the Local Volume up to 20\Mpc, stars at their centers are hard to resolve with current instrumentations. Besides, there are fewer stars brighter than the limiting magnitude, making the catalog-based searches less effective beyond 5\Mpc. Nevertheless, some of these dwarf galaxies remain visually detectable in imaging data \citep{Danieli2018}. To fully leverage the information contained in these images, we present a novel image-based search method designed to enhance the detectability of distant dwarf galaxies within the Local Volume ("LV dwarf galaxies" afterward).

A typical approach adopted in exisiting studies is to first detect extended sources from images by setting thresholds for signal to noise ratio (S/N) and angular size. Candidate sources are then validated through visual inspection \citep{Carlsten&2020}. While practical for searching satellite galaxies around individual host galaxies, this approach becomes prohibitively inefficient for large-area surveys with the CSST and future imaging survey experiments. These large datasets demand fully automated pipelines to screen candidates, as manual inspections are unsustainable. Fortunately, the distinguishing features of LV dwarf galaxies, traditionally leveraged by human classifiers to separate them from distant galaxies or galaxy groups, are inherently compatible with modern image recognition models.

Image recognition models have been widely applied to identify specific astronomical objects, such as classifications of gravitational lensing systems \citep{Shu2022} and galaxy morphologies \citep{Robertson2023,Iglesias2024}. Recent advancements in machine learning have led to the development of highly transferable models \citep{Bhavanam2024}, which achieve high performance with minimal computational effort through fine-tuning of pre-trained architectures. Among these, the Vision Transformer (ViT) -- a transformer-based model originally designed for natural image processing -- has emerged as a powerful tool. In astronomy, ViT has demonstrated superiority over traditional Convolutional Neural Networks (CNNs) across a range of tasks, including galaxy morphology classification \citep{Yao2021}, gravitational lensing detection \citep{Huang2022}, and cosmological parameter inference \citep{Gondhalekar2024}.

This work constitutes the second paper in our series on nearby dwarf galaxy detection with the CSST, extending the scope of our previous study (\citealt{qu2023}; \citetalias{qu2023} hereafter), which focused on systems within 1 Mpc. Here, we expand the search radius to 20 Mpc. The core of our approach is utilizing a transformer-based image recognition ViT model to improve the CSST's ability to detect LV dwarf galaxies over a broader distance range and to better identify faint, small systems.

This paper is organized as follows: in Section~\ref{sec:mock}, we introduce the CSST image simulation pipeline, including the fiducial input catalogs that contain artificial LV dwarf galaxies and the CSST Image Simulator. The image-based detection method, considering different background galaxies in fiducial catalogs and its performance is described in Section~\ref{sec:preprocess} and Section~\ref{sec:model}. We discuss the post-processing in Section~\ref{sect:postprocess}.
The summary is provided in Section~\ref{sec:sum}.
\begin{figure*}
   \centering
   \includegraphics[width=16cm, angle=0]{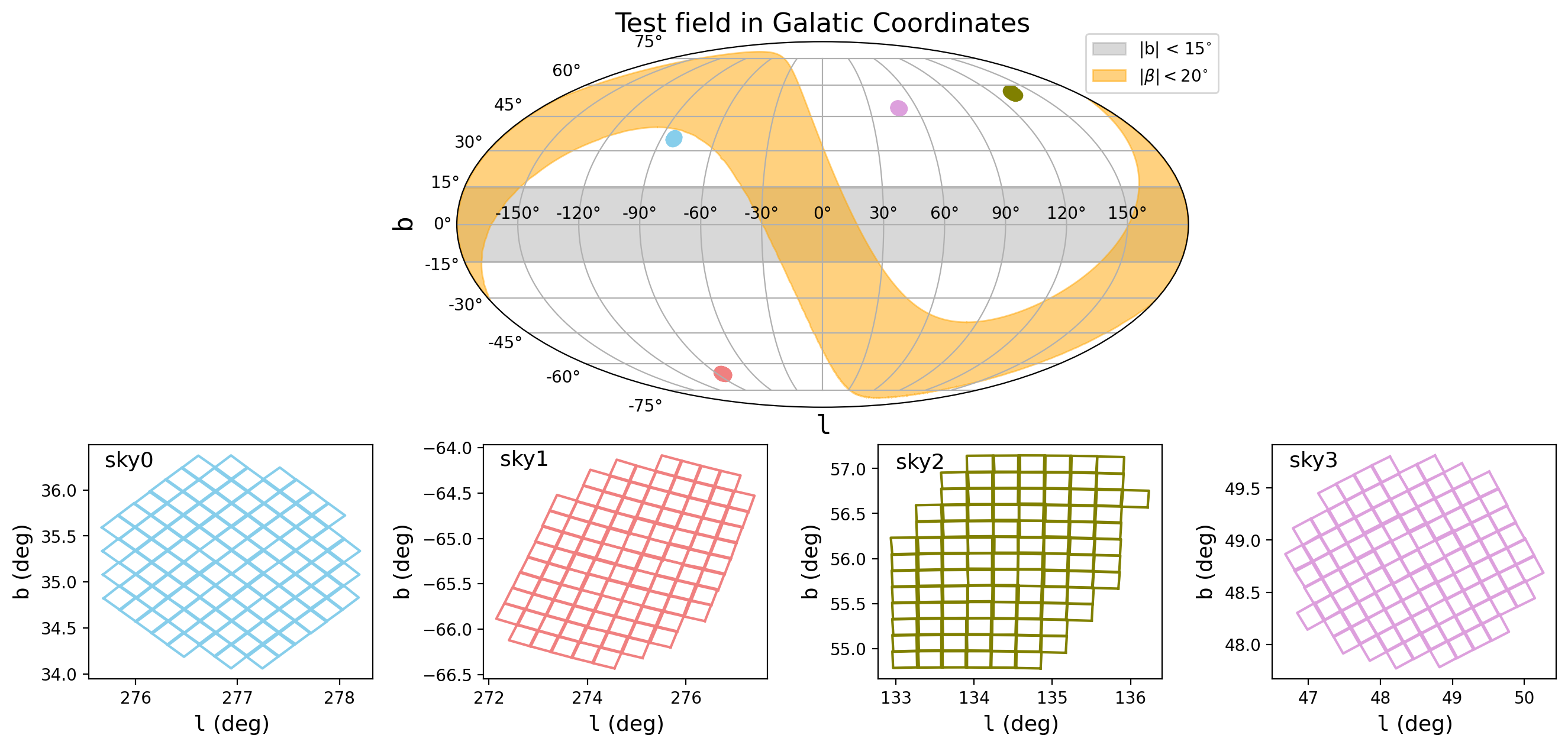}     
   \caption{Distribution of the simulated sky regions in Galactic coordinates. The gray and orange regions correspond to |b| < 15$^\circ$ and |$\beta$| < 20$^\circ$, respectively, which are areas not covered by the main CSST survey.
   }
   \label{fig:lbsky}
\end{figure*}
\begin{figure*}
   \centering
   \includegraphics[width=16.5cm, angle=0]{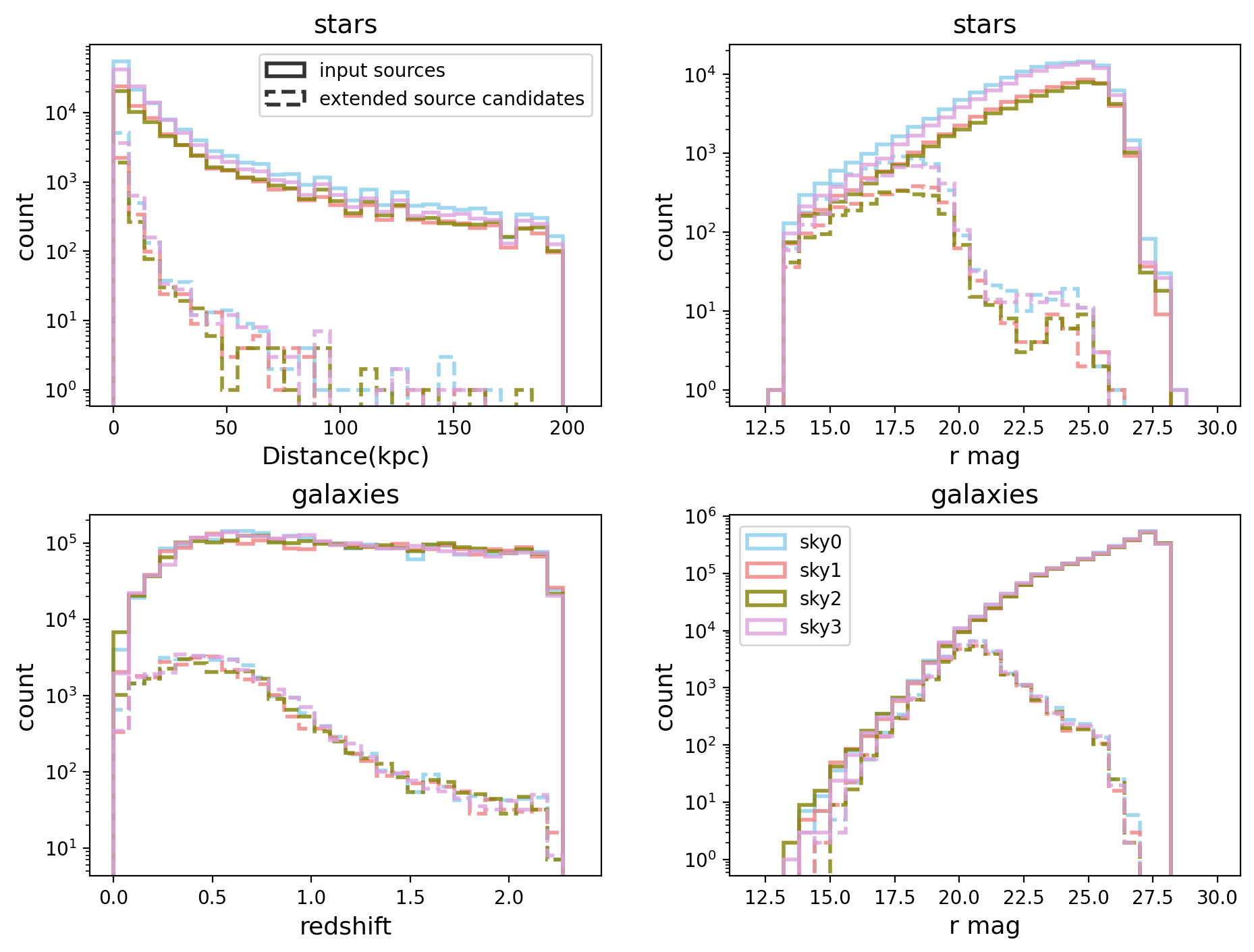}     
   \caption{The distance (redshift) and apparent magnitude(in the r band) distributions of stars and background galaxies in the fiducial catalog for the simulation program across our four test sky regions. Different colors represent data from different sky regions, with solid lines indicating the input catalog and dashed lines representing sources detected by the extended source detection algorithm in Section~\ref{subsec:esource} .
   }
   \label{fig:inputcat}
\end{figure*}
\begin{figure*}
   \centering
   \includegraphics[width=16.5cm, angle=0]{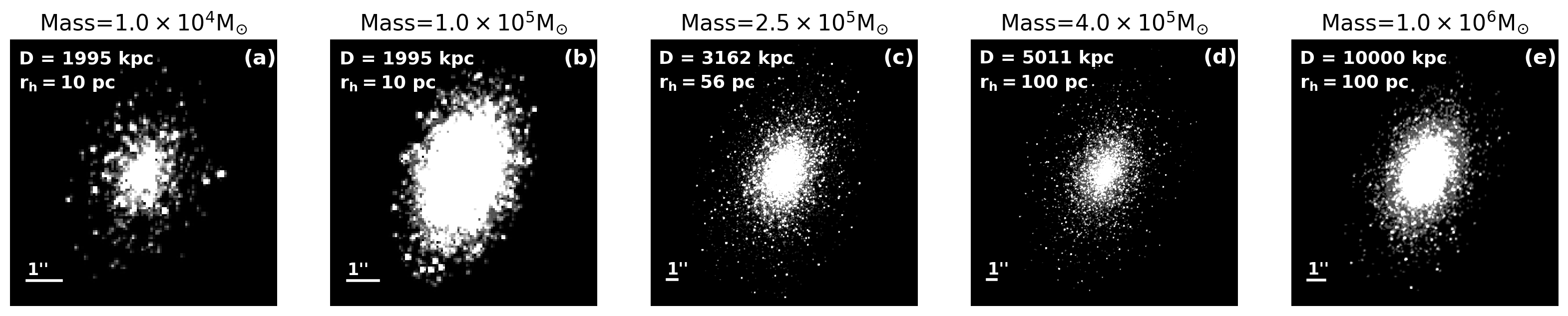}     
   \caption{Simulated images of LV dwarf galaxies in the g-band. Each panel displays a background-free image of a simulated dwarf galaxy with a stellar mass of $10^4$ -- $10^6$\msun. The stellar mass, distance, and half-light radius of each simulated galaxy are annotated in the corresponding panel.
   }
   \label{fig:dwarf}
\end{figure*}

\section{Mock Images}
\label{sec:mock}
The construction of realistic mock images as close to the real observations as possible is the first and a crucial step in our work. This process starts with a fiducial catalog that contains galaxies and stars. Then a set of artifical LV dwarf galaxies are injected into it. This mock catalog is input into the CSST Image Simulator(
Wei et al., in preparation) to produce the mock CSST images.

\subsection{Fiducial catalog}
\label{subsec:fiducial}

The fiducial catalog consists of two components: a synthetic Milky Way stellar population generated using the population synthesis model \textit{TRILEGAL} \citep{Girardi2016}, which produces mock stellar catalogs based on the Galactic structure components (thin disk, thick disk, halo, and bulge) while accounting for the extinction, photometric systems, and star formation histories; and galaxies from the cosmological simulation suite \textit{Jiutian-1G} \citep{Han2025}, referred to as background galaxies in this work (see Wei et al., in preparation). 
\textit{Jiutian-1G} is a high-resolution dark matter-only N-body cosmological simulation performed using the LGadget-3 code, adopting the Planck 2018 cosmological parameters \citep{Planck2020}. The simulation spans a comoving box of 1000 $h^{-1}$ Mpc per side with $6144^3$ particles and achieves a mass resolution of $3.723 \times 10^8 M_\odot$. Galaxies are populated using both semi-analytical models (e.g., GAEA, LGalaxies) and subhalo abundance matching (SHAM).
Based on this simulation, a mock light-cone galaxy catalogue extending to redshift $z \sim 3.5$ is constructed. A full-sky ray-tracing simulation is conducted to obtain weak gravitational lensing signals, including shear and magnification, at each galaxy's position. The final galaxy catalogue includes positions, redshifts, stellar masses, morphologies, sizes, spectral energy distributions (SEDs), shear, and magnification.

As shown in Figure~\ref{fig:lbsky}, our fiducial catalog comes from four sky regions, which represent typical CSST regions by avoiding $|b|$ $<$ 15$^{\circ}$ (galactic latitude) and $|\beta|$ $<$ 20$^{\circ}$ (ecliptic latitude). Every sky region includes 100 mosaicking patches that corresponds to the projected CCD footprint of a single chip covering $11\farcm4 \times 11\farcm4$ on the sky. We generate source catalogs in three photometric bands: $g$, $r$, and $i$. The distance distribution of the Milky Way stars as well as their luminosity function in the $r$-band magnitudes of each sky region is shown as solid histograms in the upper row of Figure~\ref{fig:inputcat}. Similarly, the redshift distribution and the luminosity function of the background galaxies are shown in the bottom row. The faint end of the luminosity function of both stars and galaxies are well below the depth ($r$ $\approx$ 25.5mag) of the CSST survey for a single exposure of 150s.

\subsection{Artificial Local Volume dwarf galaxies}
\label{subsec:dwarfs}

\renewcommand{\arraystretch}{1.2}
\begin{table} 
	\caption{Parameters of artificial LV dwarf galaxies.}
	\setlength{\tabcolsep}{8pt}
	\label{tab:mockdata}
	\centering  
	\begin{tabular}{l|lll} 
		\hline
		Parameter& Minimal & Maximal  &Step (log scale)\\
		\hline
		Stellar Mass (\msun)     & $10^{3}$  & $10^{6}$    & $10^{0.2}$ \\
        $r_{\rm h}$ (pc)     & 10  & 316    & $10^{0.5}$\\
        D (kpc) & 316 & 19952    &   $10^{0.25}$ \\
		\hline
	\end{tabular}
	\setlength{\tabcolsep}{8pt}
\end{table}

Following the recipe presented in \citetalias{qu2023}, artificial LV dwarf galaxies are constructed using single stellar population PARSEC models\footnote{\url{http://stev.oapd.inaf.it/cgi-bin/cmd}} \citep{Bressan2012}. Given that almost all known Local Group dwarf galaxies are predominantly old and metal-poor, we adopt a uniform age of 11 Gyr and metallicity of [M/H] = $-$2.0. The stellar radial density profiles are modeled with an exponential profile. The LV dwarf galaxies have stellar mass distributions from $10^3$\msun to $10^6$\msun and half-light radii from 10\pc to 316\pc. The distance range is set from 316\kpc to 20\Mpc, which extends the exploration from the Local Group to the Local Volume. Given that currently known LV dwarf galaxies within the Local Group are almostly complete with stellar masses above $10^5$\msun \citep{Drlica-Wagner2021,am2022pandas}, we set a lower distance limit of 1\Mpc when testing these relatively massive systems. In total, 1,953 artificial LV dwarf galaxies are generated.

The stellar catalog for each LV dwarf galaxies is constructed using the catalog of its member stars, enabling a realistic simulation of the light emission processes of nearby dwarf galaxies.

\subsection{CSST Image Simulator}
\label{subsec:simulator}

Mock images are generated using the CSST Image Simulator\footnote{\url{https://csst-tb.bao.ac.cn/code/csst_sim/csst-simulation}}, which is developed by the CSST scientific data processing and analysis system. By combining the mock galaxy catalog from cosmological simulations with weak gravitational lensing effects and detailed instrument modeling, the simulator is aimed to produce realistic mock images with comprehensive instrumental and observational features.

To accurately model the impact of the optical system on image quality, a comprehensive simulation model of the CSST optical system has been developed to generate high-fidelity point spread functions (PSFs). This optical simulator consists of six distinct modules that account for various optical aberrations, including mirror surface roughness, fabrication imperfections, CCD assembly errors, and thermal-induced distortions. Additionally, the simulator incorporates two dynamic error sources, micro-vibrations and image stabilization effects, providing a realistic representation of the PSF under operational conditions.

To produce realistic mock images, various noises have been included, such as shot noise, sky background noise, and detector-related effects. Using the throughput system of CSST, photons from each galaxy are generated with Galsim\footnote{\url{https://github.com/GalSim-developers/GalSim}} \citep{Rowe2015}. Here, the throughput system accounts for mirror efficiency, filter transmission, and the detector's quantum efficiency, ensuring that the simulated images closely match the actual observational conditions.

Poisson noise is included to model contributions from both the sky background and the CCD detector's dark current. Specifically, the i-band background level was set to 0.212 ${\rm e}^-$/pixel/s, with a dark current of 0.02 ${\rm e}^-$/pixel/s. For a 150s exposure, this results in an average signal of approximately 35 $\rm e^-$/pixel. Additionally, read noise was modeled as a Gaussian distribution with a standard deviation of approximately 5.0 ${\rm e}^-$/pixel. To simulate the generation of mock galaxy images on the detector, bias effects were included, and the gain factor was applied for calibration.

We first generate mock images based on the fiducial catalog, incorporating basic instrumental effects except for cosmic rays, hot pixels, bad columns and Charge Transfer Inefficiency (CTI). These images, referred to as "fiducial images". To facilitate flexible testing of LV dwarf galaxies detection, mock images of LV dwarf galaxies are generated independently. They omit bias, dark current, and sky background. By treating LV dwarf galaxies images as background-free overlays, this enabels us to place them freely into the fiducial images. Figure~\ref{fig:dwarf} shows examples of simulated LV dwarf galaxy images in the g-band, presented without background and annotated with key parameters, including stellar mass, distance, and half-light radius. We then inject each LV dwarf galaxy into the fiducial images 100 times by varying their on-sky projections.

\section{Pre-process}
\label{sec:preprocess}

LV dwarf galaxy searches start with an automated process of detecting extended sources. This procedure is implemented to the fiducial images that contain articial LV dwarf galaxy images. Based on the detected sources from the background and from the injected LV dwarf galaxies, we built negative and positive samples to train the AI classifier (see Section~\ref{sec:model}). 
The full implementation of both algorithmic steps is publicly available\footnote{\url{https://github.com/nemoqh77/LVdgdetection/tree/main}}.
Here, we outline the extended source detection process, which follows a similar approach to that of \citet{Carlsten&2020} and \citet{Davis&2021}.

\subsection{Extended source detection}
\label{subsec:esource}

\begin{figure*}
   \centering
   \includegraphics[width=15.5cm, angle=0]{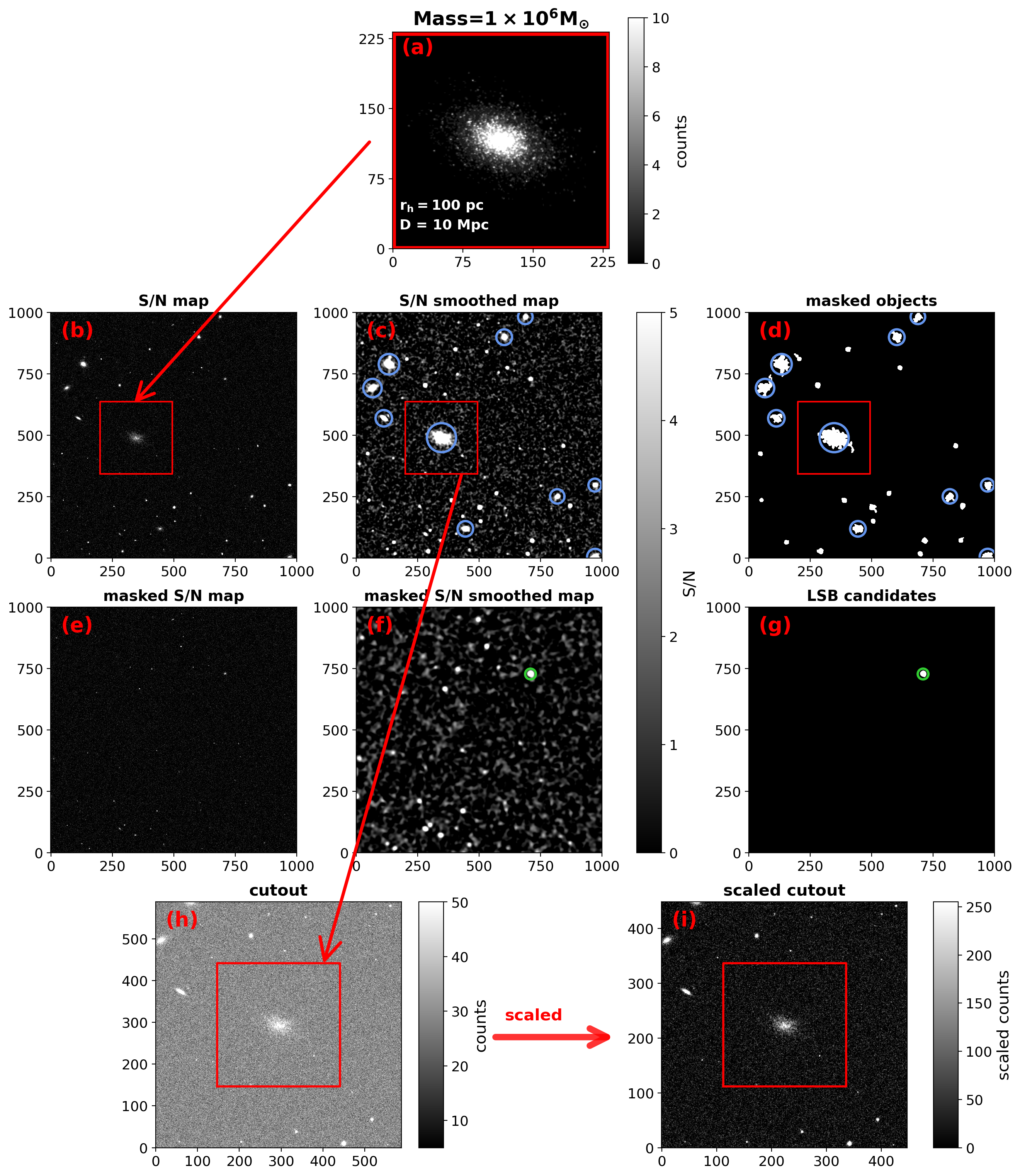}     
   \caption{Extended source detection process.
    (a) Background-free image of a simulated LV dwarf galaxy with a stellar mass of $10^{6}\,\mathrm{M}_\odot$, a half-light radius of 100 pc, and a distance of 10 Mpc.  
    (b) The same galaxy injected into a "fiducial Image" showing the S/N map obtained by dividing the observational image by the background noise. The red box marks the galaxy's location.  
    (c) S/N map from (b) after convolution with a 3-pixel kernel.  
    (d) Sources satisfying the masking threshold; blue circles highlight those meeting the high-surface-brightness (HSB) candidate criteria. The LV dwarf galaxy satisfies the selection in this case.  
    (e) S/N map after applying mask.  
    (f) Smoothed version of (e) using a 6-pixel convolution kernel.  
    (g) Detected sources satisfying the low-surface-brightness (LSB) candidate criteria.  
    (h) Cutout image centered on the detected LV dwarf galaxy.  
    (i) Scaled version of (h), with a red box indicating the central $224 \times 224$ pixel region, covering one-quarter of the cutout area.  
    A unified colorbar is used in panels (b), (c), (e), and (f) to indicate the S/N values, while colorbars in panels (a), (h), and (i) represent flux intensity.}
   \label{fig:detection}
\end{figure*}

\begin{figure*}
   \centering
   \includegraphics[width=16.5cm, angle=0]{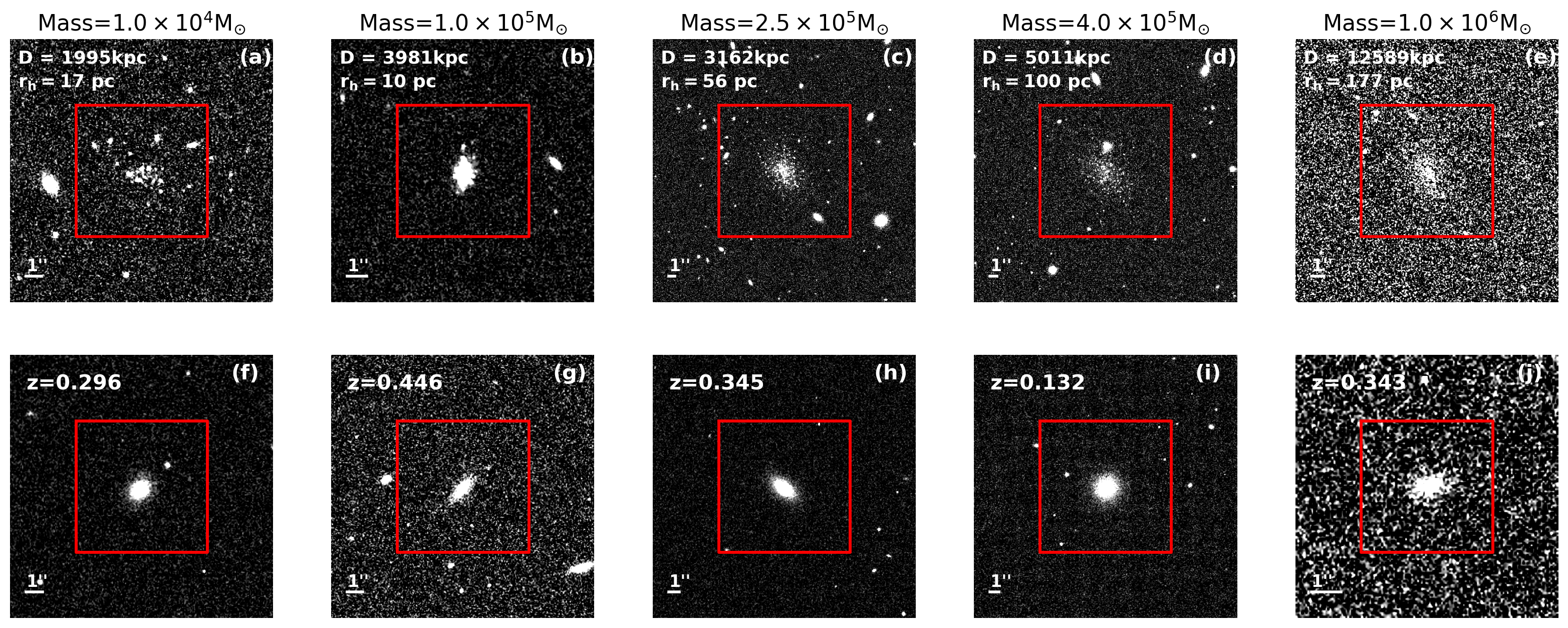}     
   \caption{Scaled cutouts of extended source candidates in the $g$-band, obtained from the extended source detection process. The first row shows true LV dwarf galaxies. The second row shows negative samples detected in the fiducial images; these are primarily distant galaxies, with their respective redshifts indicated in the image annotations.
   }
   \label{fig:show_candidates}
\end{figure*}

The extended source detection process is applied to the mock images, where high-surface-brightness (HSB) systems are first identified. Masking these bright sources then facilitates the detection of low-surface-brightness (LSB) systems. We divide each mock image (9232$\times$9216 pixels) into small areas of 1000 $\times$1000 pixels, with an overlap of 100 pixels between adjacent areas. The detection procedure is performed in each area as outlined below and also illustrated in Figure~\ref{fig:detection}:
\begin{enumerate}[label=\arabic*.,leftmargin=*]
    \item Generate the S/N map shown as panel (b).
    \item Produce the smoothed $S/N$ map using gaussian convolution with a 3$\times$3-pixels kernel shown in panel (c), and select regions around a given candidate source using $S/N$ thresholds. 
    
    In our practice, two regions are selected: $S_1$ with $S/N$ $>$6 and $S_2$ with $S/N$ $>$1.5. The sources satisfy all three criteria below are classified as HSB candidates higlighted by the blue cricles in panels (c) and (d);
    \begin{enumerate}[label=\roman*.,labelsep=1em,leftmargin=*]
        \item $S_1$ $>$ 60 pixel$^2$
        \item $S_2$ $>$ 600 pixel$^2$
        \item $S_1$ $/$ $S_2$ $>$ 0.2
    \end{enumerate}
    The sources satisfied only criteria (i) and (iii) are classified as bright source contaminants. These typically correspond to bright stars or compact galaxies that do not exhibit the characteristic profiles expected for LV dwarf galaxy candidates. These contaminants are shown by the white regions without overlaid blue circles in panel (d).
    \item Mask the S/N map from (b) with bright sources (HSB and contaminants) in panel (d). Replace the masked region with local background noise to create the masked $S/N$ map in panel (e).
    \item Produce the smoothed $S/N$ map after masking (e) with a Gaussian convolution with a 6$\times$6-pixels kernel in panel (f).
    \item Detect LSB candidates by selecting region $S_2$ with $S/N$ $>$ 4 and $S_2$ $>$ $400$ pixel$^2$, which are highlighted with green cricles in panel (f) and (g).
    \item Extract images around HSB and LSB candidates by selecting regions of 4$\times S_2$ shown as the red box in panel (h). Cross-match the overlapping areas to eliminate redundant detections.
\end{enumerate}

The above procedure is performed independently in the $g$, $r$, and $i$ bands. The results from these three bands are then cross-matched. Only detections with a tolerance of less than 1" are retained as the final detected extended sources.

We identify the extended source candidates from all four sky regions. For all the sources, we find the corresponding sources in the input catalog by performing cross-match with 1" radius. This ensures that the detection procedure is finding "true" systems. 

Overall, about 3\% of the input sources are identified, the vast majority of which are stars and galaxies brighter than $r$ = 20, represented by the dashed histograms in Figure~\ref{fig:inputcat}. There are about 300 extended sources detected from a skypatch of $11\farcm4\times11\farcm4$, yielding around 24,000 sources from thr first region, sky0 (see Figure~\ref{fig:lbsky}). These detected sources are used to construct the negative samples for the image-classifier.
Using the same extended source detection process, detection tests are conducted on 1,953 synthetic LV dwarf galaxies, each injected 100 times with random rotations into fiducial images from sky0, with their individual detection rates summarized in Figure~\ref{fig:step1_detectivity}. This procedure yields 62,689 cutouts from 1,143 successfully recovered galaxies, which serves as the basis for constructing the positive sample set.

Figure~\ref{fig:show_candidates} presents scaled cutouts of extended source candidates in the g-band. The first and second rows of panels correspond to LV dwarf galaxies and other candidates, respectively. The latter are detected in the fiducial images and are primarily distant galaxies. A clear morphological distinction is observed between the two categories: LV dwarf galaxies often display partially resolved structures with discernible outlines of individual member stars, while distant galaxies appear as unresolved sources lacking visible stellar features. These morphological distinctions form the basis for employing machine learning techniques to classify and separate the two types of sources effectively.

Extrapolating the number of sources from a single skypatch to the full CSST survey footprint of 17,500 square degrees, we anticipate detecting over tens of millions of extended sources as contaminants. Given the enormous number of sources, traditional visual inspection to identify LV dwarf galaxies is almost impossible, and thus an automatic image-based image recognition method is necessary for this task.

\begin{figure*}
   \centering
   \includegraphics[width=16.5cm, angle=0]{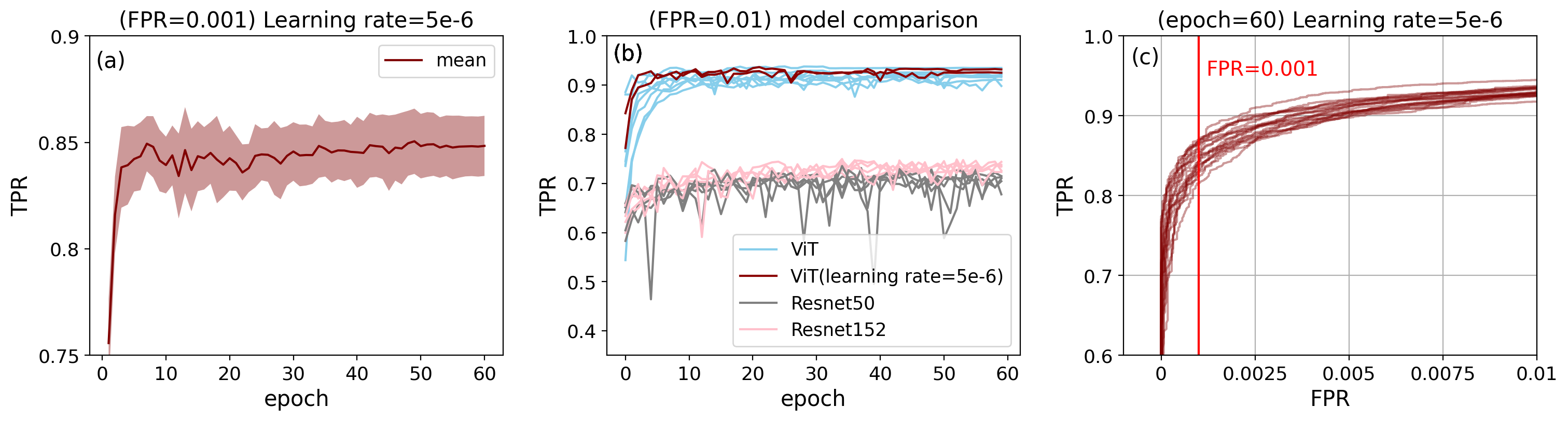}     
   \caption{Performance evaluation of different models and training configurations based on true positive rate (TPR).Panel (a): Evolution of TPR during training at FPR = 0.001. The red solid line indicates the mean TPR across 20 dataset samples, and the shaded region denotes the 1$\sigma$ deviation.  Panel (b): Comparison of training performance among ResNet50, ResNet152, and ViT models, trained on linearly scaled sky0 data using $N_{\rm max} = 15$ (Set 1, Group A). The x-axis represents training epochs, and the y-axis shows the TPR at FPR = 0.01. Pink, gray, and blue lines correspond to ResNet50, ResNet152, and ViT, respectively. Panel (c): Classification performance of ViT at epoch 60. Each curve shows the TPR-FPR relation for one of 20 independent testing set samples.}
   \label{fig:set10_tf_vit}
\end{figure*}

\subsection{Positive and negative samples}
\label{subsec:samples}

To prepare the construction of the positive and negative samples for the image-classifier, we first standardize the size of the input images to a uniform resolution of 446 $\times$ 446 pixels in 3 channels which correspond to the three bands. We then normalize the flux of the image using linear scaling shown in panel (i) of Figure~\ref{fig:detection}. In order to make the training process more efficient, we zoom into the central 1/4 region of each image, yielding the final image size of 224$\times$224. 

To construct the positive training dataset, we begin with 1,143 LV dwarf galaxies that are successfully detected at least once during the extended source detection stage. Each galaxy is tested through 100 synthetic placements followed by detection, with each successful detection yielding an image cutout. The total number of cutouts per LV dwarf galaxy varies according to its detection rate in the extended source detection process. To prevent bright or easily detectable galaxies from dominating the dataset, an upper limit of $N_{\rm max} = 15$ cutouts per LV dwarf galaxy is applied. Galaxies with fewer than $N_{\rm max}$ detections contribute all their available cutouts, ensuring a balanced distribution across systems of varying detectability. For comparison, we also perform experiments with $N_{\rm max}$ values of 30, 50, 75, and 100 (see Section~\ref{subsec:effects} for details).

From these 1,143 LV dwarf galaxies, we randomly select 300 galaxies (with their associated cutouts) as the training set, 200 as the validation set, and assign the remaining 643 galaxies to the testing set. These three subsets are mutually exclusive: each LV dwarf galaxy (along with its associated cutouts) is assigned to only one group. This ensures that no galaxy appears in more than one subset, so the model's performance on the testing set reflects genuine generalization. We refer to this configuration as Group A. In this setting, over half of the LV dwarf galaxies are assigned to the testing set. Then we construct a corresponding Group B for each Group A configuration. Specifically, from the 643 LV dwarf galaxies in the Group A testing set, we randomly select 300 for training and 200 for validation in Group B. The remaining 143, combined with the 500 LV dwarf galaxies previously used in Group A's training and validation sets, form Group B's testing set. Group A and Group B are used independently for model training and evaluation. This design ensures that all LV dwarf galaxies appear in at least one testing set across Groups A and B, thereby enabling us to assess the ViT model's classification performance for each LV dwarf galaxy.

Negative samples are drawn from extended sources detected in the fiducial images, with  $\sim$ 30,000 sources in each sky region (as shown in Figure~\ref{fig:lbsky}).
In sky0, which serves as the primary region for training and validation, a total of 29,152 cutout images are obtained from the extended source detection pipeline. To ensure class balance, we randomly select 3,200 and 2,200 samples as negative examples for the training and validation sets, respectively, matching the size of the corresponding positive sets. The remaining 23,752 samples are reserved as the negative component of the testing sets. To further assess model generalization, we construct three additional testing sets using extended source detections from the remaining sky regions (sky1, sky2, and sky3), which contain 27,298, 26,255, and 32,079 cutouts, respectively. 

This sampling strategy is applied independently for each of the 10 Group A/B dataset pairs constructed under the $N_{\rm max} = 15$ configuration, yielding 20 distinct datasets. Table~\ref{tab:dataset} summarizes the sample sizes for Group A. For each dataset, the negative training and validation samples are randomly drawn from sky0. The remaining cutouts from sky0-after removing those selected for training and validation-constitute the testing set for that specific dataset. As a result, the exact composition of the negative testing sets differs across the 20 dataset pairs.

\renewcommand{\arraystretch}{1.2}
\begin{table} 
	\caption{Group A of ten sets in sky0}
	\setlength{\tabcolsep}{8pt}
	\label{tab:dataset}
	\centering  
	\begin{tabular}{l|lll} 
		\hline
		positive / negative & training set  & validation set  & testing set\\
		\hline
		Set 1 & 3152/3200 & 2088/2200 & 6929/23752\\
        Set 2 & 3290/3200 & 2136/2200 & 6743/23752\\
        Set 3 & 3152/3200 & 2000/2200 & 7017/23752\\
        Set 4 & 3195/3200 & 2251/2200 & 6723/23752\\
        Set 5 & 3176/3200 & 2118/2200 & 6875/23752\\
        Set 6 & 3262/3200 & 2018/2200 & 6889/23752\\
        Set 7 & 3242/3200 & 2216/2200 & 6711/23752\\
        Set 8 & 3274/3200 & 2129/2200 & 6766/23752\\
        Set 9 & 3049/3200 & 2255/2200 & 6865/23752\\
        Set 10 & 3189/3200 & 2197/2200 & 6783/23752\\
		\hline
	\end{tabular}
	\setlength{\tabcolsep}{8pt}
\end{table}

\section{Image Classification}
\label{sec:model}
\subsection{ViT model}
\label{subsec:model}

The identification of LV dwarf galaxies can be translated to binary classification task. The model used for this task is the Vision Transformer (ViT) \citep{vit2020}, an advanced architecture based on the Transformer framework. The ViT model divides input images into patches of fixed size, and processes them as tokens through a self-attention mechanism. 
In contrast to CNNs that rely on convolutional operations to extract local features \citep{cnn2017}, the ViT model uses a self-attention mechanism to establish relationships across the entire image, making it highly effective at capturing global contextual information.
This advantage is particularly beneficial for tasks like galaxy classification, where understanding spatial and structural relationships is critical. By pre-training on large datasets such as ImageNet-21k, a dataset containing over 21,000 categories of images, the ViT learns transferable features that can be fine-tuned for specific astronomical applications, ensuring robust performance even in the presence of variations in image quality, resolution, and noise.

In our implementation, we train the "vit-base-patch16-224-in21k" model using the candidate cutouts obtained from Section~\ref{subsec:samples}. The ViT variant is pre-trained on the ImageNet-21k dataset, which contains 14 million images across 21,000 categories, providing a robust foundation for transfer learning.

\begin{figure*}
   \centering
   \includegraphics[width=17.5cm, angle=0]{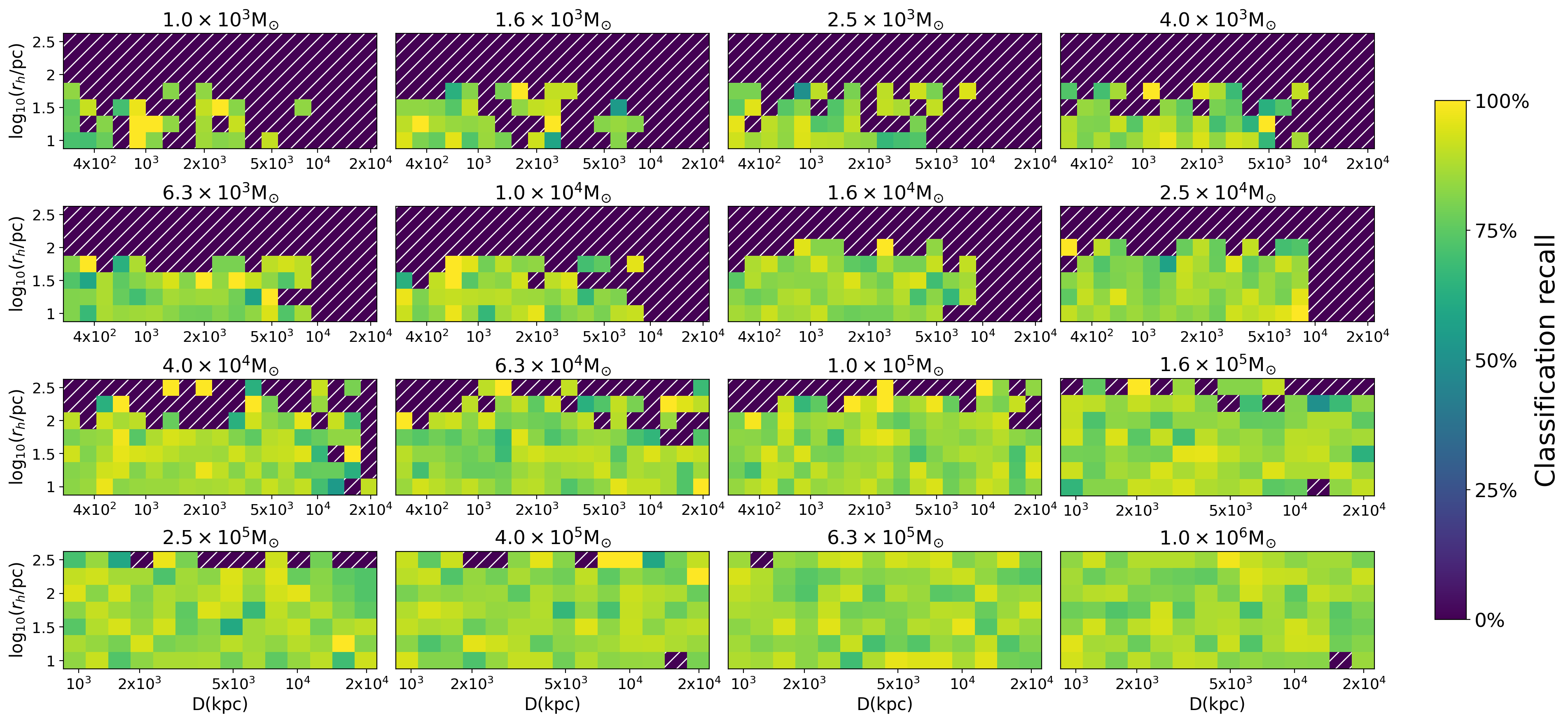}     
   \caption{Mean classification recall of the ViT model for LV dwarf galaxies across all 20 datasets. The color of each pixel corresponds to the TPR at FPR = 0.001 for the respective LV dwarf galaxies. The shaded regions indicate areas without test samples.}
   \label{fig:vitdetection}
\end{figure*}

\begin{figure*}
   \centering
   \includegraphics[width=16.5cm, angle=0]{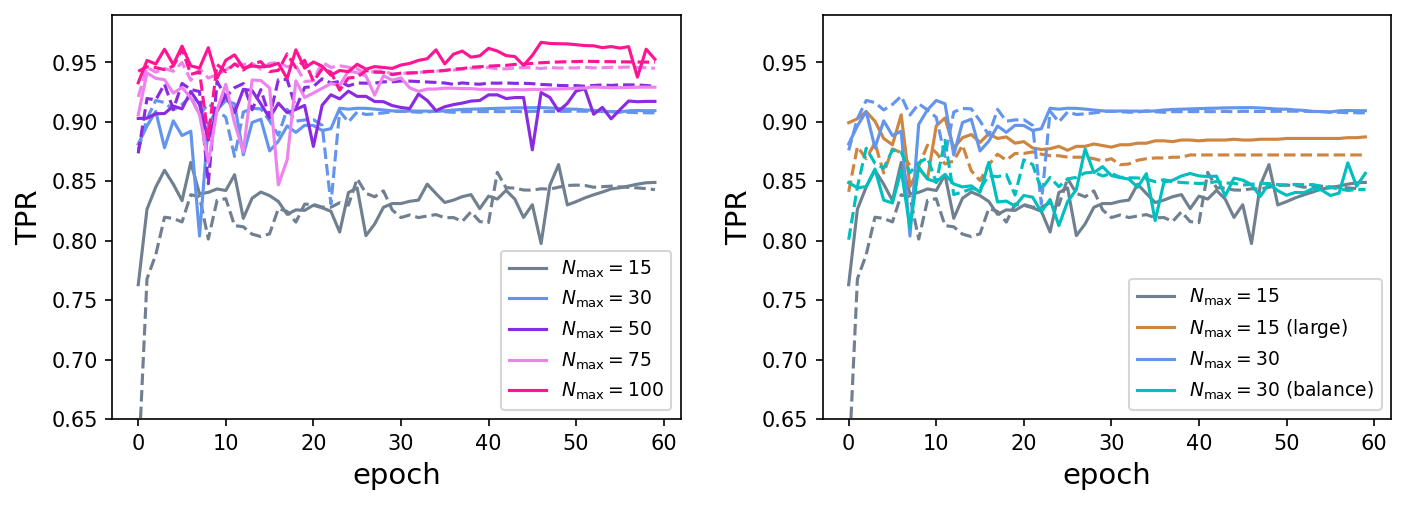}     

   \caption{ViT classification recall under different datasets.The x-axis represents the number of training epochs, and the y-axis corresponds to the TPR at FPR = 0.001. Lines in different colors represent results for different $N_{\rm max}$ values or datasets. For lines of the same color, solid and dashed lines correspond to the results of samples A and B in Set 1.}
   \label{fig:dataset_compare}
\end{figure*}

\begin{figure*}
   \centering
   \includegraphics[width=17.5cm, angle=0]{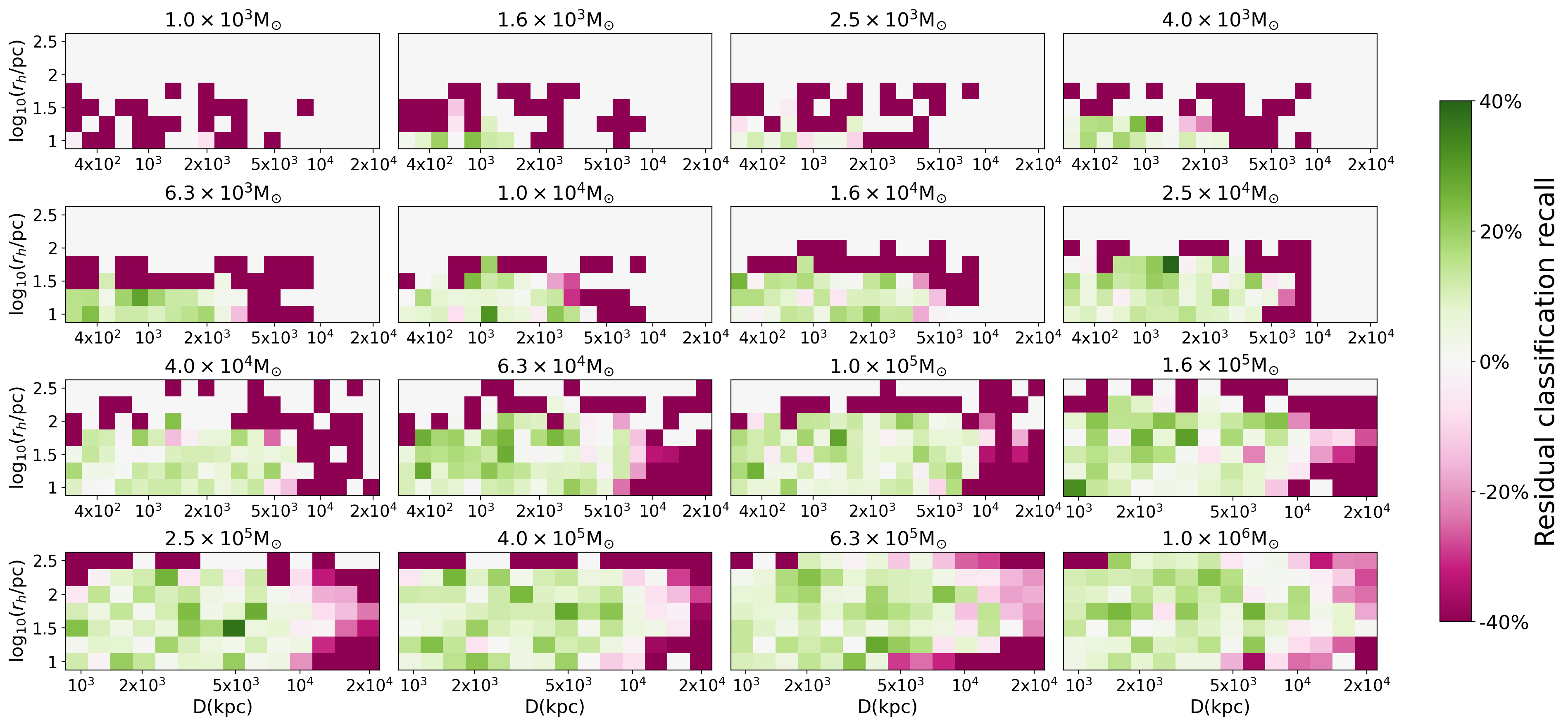}     
   \caption{Difference map of ViT classification recall between $N_{\rm max}$ =30 and $N_{\rm max}$ =15 across 20 datasets. The figure shows the difference in TPR distributions (at FPR = 0.001) between $N_{\rm max}$ =30 and $N_{\rm max}$ =15. Pink pixels indicate LV dwarf galaxies for which the TPR in the $N_{\rm max}$ =30 group is lower than that in the $N_{\rm max}$ =15 group.}
   \label{fig:residual}
\end{figure*}

\subsection{Classification results}
\label{sec:results}

We evaluate the classification performance of the Vision Transformer (ViT) model on the testing sets, following training and validation using the datasets described in Section~\ref{subsec:samples}. We trained ViT models on all 20 datasets use a learning rate of $5 \times 10^{-6}$, and the results are summarized in Figure~\ref{fig:set10_tf_vit}. The false positive rate (FPR) is fixed to 0.001 to ensure a low contamination level. The left panel of Figure~\ref{fig:set10_tf_vit} displays the learning curve of the model, showing the true positive rate (TPR) as a function of training epochs. The red solid line denotes the mean TPR, and the shaded region represents the 1$\sigma$ uncertainty. The model converges to a mean TPR of approximately 85\% at around the 60th epoch. The middle panel illustrates the learning curves for different learning rates, evaluated on the Set 1 dataset at FPR = 0.01. The middle panel of Figure~\ref{fig:set10_tf_vit} presents the learning curves under various learning rates, evaluated on the Set 1 dataset at a fixed FPR of 0.01. For comparison, we also include the performance of two CNNs, represented by the grey and pink lines, which converge to a TPR of approximately 70\%. In contrast, the ViT model, shown as the blue and dark red curves, achieves significantly higher TPRs of around 90\%, demonstrating superior classification performance under the same FPR constraint. More comparisons of different learning rates are shown in Figure~\ref{fig:tpr_vit}.
The right panel shows the TPR-FPR relation after 60 training epochs, with each line representing a different dataset. The consistency across datasets highlights the stability and robustness of the ViT model. In addition, based on tests using the Set 1 data, we observe no significant difference in model predictions across the four sky regions (sky0 to sky3) as shown in Figure~\ref{fig:tpr_vit}, further supporting the model's generalization capability.

To further evaluate classification accuracy on a per-galaxy basis, we aggregate the results from all 20 datasets. We define the classification recall of a given LV dwarf galaxy as the TPR at a fixed FPR of 0.001, as illustrated in Figure~\ref{fig:vitdetection}. Each panel displays the classification recall for LV dwarf galaxies of the same stellar mass, plotted as a function of distance (x-axis) and half-light radius (y-axis), with color encoding the classification recall. The results indicate that detection rates are comparable across galaxies with different morphological properties, demonstrating that the ViT model effectively generalizes over key physical properties of LV dwarf galaxies.

\subsection{Effects of imbalanced datasets}
\label{subsec:effects}

In supervised learning, the training results depend heavily on how the datasets are constructed. In this work, there are two key parameters in constructing the datasets that would have an impact on the training: the selection of $N_{\rm max}$ and the number of artificial LV dwarf galaxies used to draw training and validation sets (see Section~\ref{subsec:samples}). 

In the baseline configuration ($N_{\rm max}=15$), we demonstrated that per-galaxy sampling caps effectively mitigate sample imbalance and lead to comparable classification recall across different LV dwarf galaxies (Figure~\ref{fig:vitdetection}). To further quantify the impact of $N_{\rm max}$, we explored larger caps. As shown in Figure~\ref{fig:dataset_compare}, the completeness steadily increases with $N_{\rm max}$, reaching TPR $\gtrsim 90\%$ for $N_{\rm max} \ge 30$. However, a closer look reveals that this gain is driven primarily by brighter, nearby galaxies, while fainter systems suffer reduced classification recall when high-detection galaxies dominate the training set (Figure~\ref{fig:residual}, pink bins).

To mitigate this imbalance, we generated an augmented dataset by increasing the number of detection tests for galaxies with pre-process detection rates below 30\%, adding 200 additional cutouts per source. This balanced configuration, denoted as $N_{\rm max} = 30$ (balance), yields performance comparable to the baseline $N_{\rm max} = 15$ setup (green line in right panel of Figure~\ref{fig:dataset_compare}), with a TPR of $\sim$85\% at FPR = 0.001. These results highlight that, although larger $N_{\rm max}$ values can improve completeness, careful balancing is required to maintain sensitivity to faint systems and avoid bias toward bright galaxies. 

The sampling strategy is another key factor influencing model performance. In the fiducial configuration, we draw $\sim$3,200 cutouts from 300 LV dwarf galaxies for training, $\sim$2,200 cutouts from 200 galaxies for validation, and assign the remaining 643 galaxies to the testing set (see Section\ref{subsec:samples}). This setup allocates more than half of the LV dwarf galaxies to evaluation, providing a stringent test of the ViT model's classification ability across a wide variety of galaxies, assuming that morphological similarities within the parameter space of dwarf galaxies allow for effective generalization by the model. However, such a strict configuration limits the training diversity available to the model. To investigate the effect of a more training-heavy sample split, we construct a control setup, referred to as "$N_{\rm max} = 15$ (large)", in which the training and validation sets are expanded to include 600 and 400 LV dwarf galaxies, respectively, leaving 143 for testing. As shown in the right panel of Figure~\ref{fig:dataset_compare}, this setup leads to a TPR exceeding 87\% on the testing set (brown-yellow line). The result confirms that increasing the number of training examples improves the model's ability to generalize, though at the cost of a smaller testing set.

Building on the previous results, we find that increasing the diversity and quantity of LV dwarf galaxies in the training set improves classification performance. However, excessively large training sets, especially those derived from synthetic data may lead to overfitting, potentially limiting model generalizability when applied to real observations. Because our current dataset is based on mock images, a more rigorous evaluation will require testing against real observational data. While archival images of known LV dwarf galaxies are available from existing surveys, their quality and characteristics are highly dependent on the real instrumentation performance, particularly the optical resolution. A more robust assessment will thus only become feasible once CSST survey data become available.

The primary goal of this work is to establish and validate a complete detection pipeline using controlled simulations. Once CSST observations commence, the model will need to handle additional complexities such as crowded stellar fields, proximity to large galaxies, and background contamination from diffuse light. These environments can significantly affect detection reliability. Future adaptations may also involve tuning extended source detection thresholds and retraining on hybrid datasets combining mock and real observations to enhance robustness.

 \begin{figure*}
   \centering
   \includegraphics[width=15.5cm, angle=0]{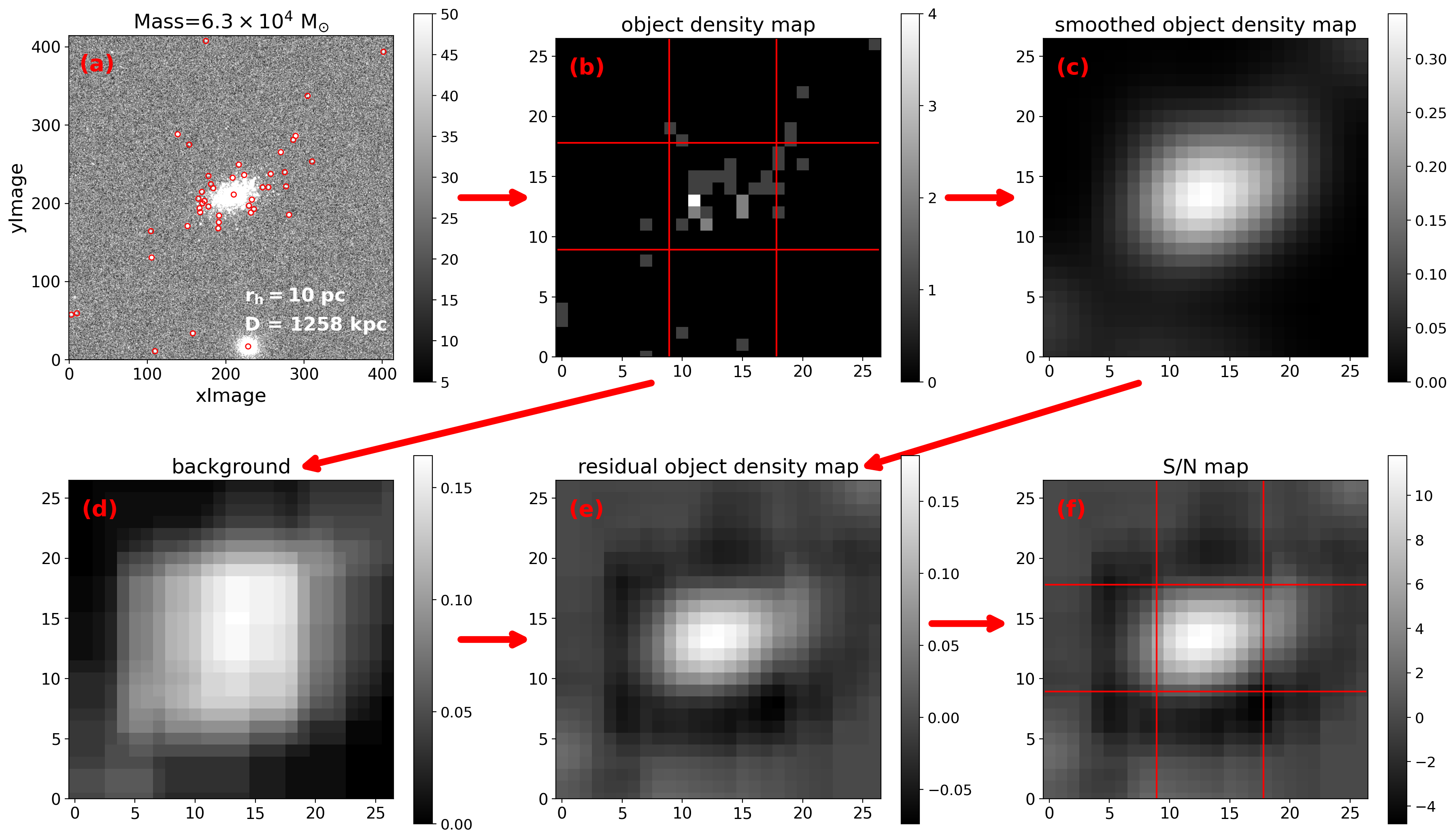}    
   \caption{Post-Process Workflow Illustration. (a) Cutout image with red circles marking the objects detected by the Source Extractor. (b) Density map of objects corresponding to (a), with each pixel representing 1". (c) and (d) are the results of convolving (b) with kernels of "$\sigma$ = 1" and "$\sigma$ = 28," respectively. (e) Residual density map obtained by subtracting (d) from (c). (f) S/N map derived by dividing (e) by the background noise.
   }
   \label{fig:postprocess}
\end{figure*}

\begin{figure*}
   \centering
   \includegraphics[width=15.5cm, angle=0]{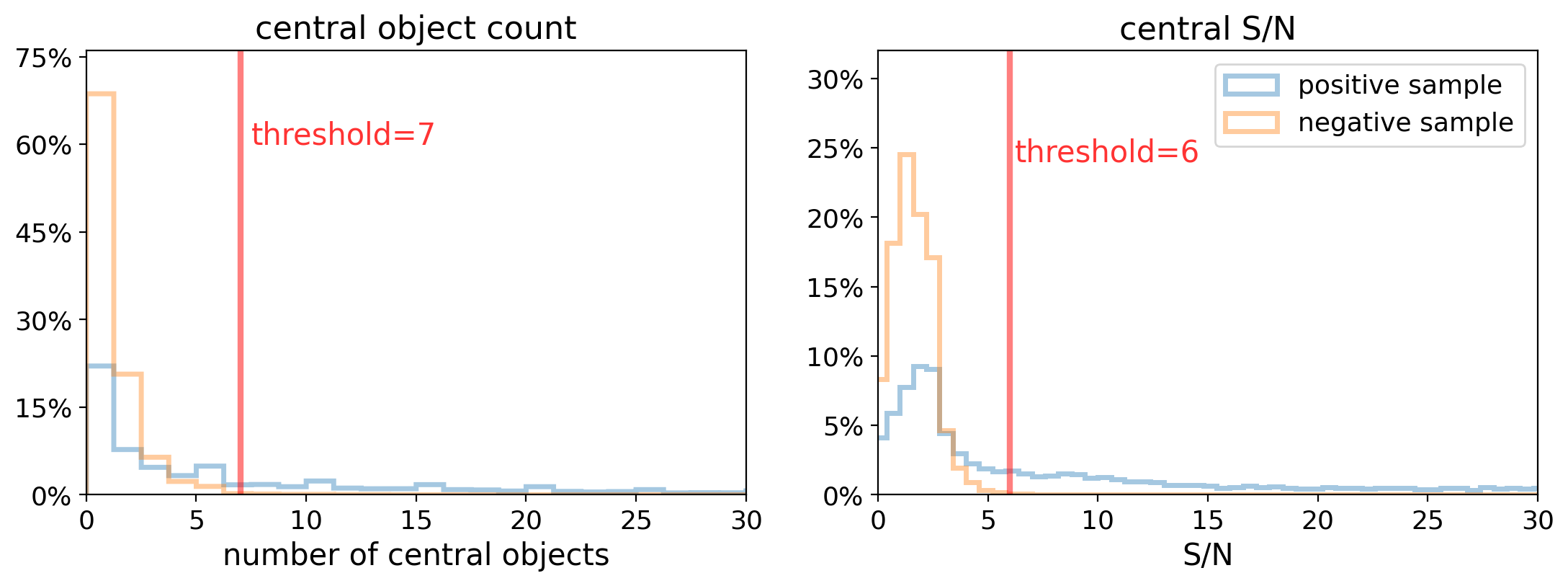}     
   \caption{Distribution of "central object count" and "central S/N" for all samples classified as negative by the ViT model, including true positive samples (blue line) and true negative samples (orange line). The vertical axis of the histogram represents the percentage of samples. Red lines indicate the thresholds used for LV dwarf galaxy selection in the post-process step.
   }
   \label{fig:sn_postprocess}
\end{figure*}

\begin{figure*}
   \centering
   \includegraphics[width=17.5cm, angle=0]{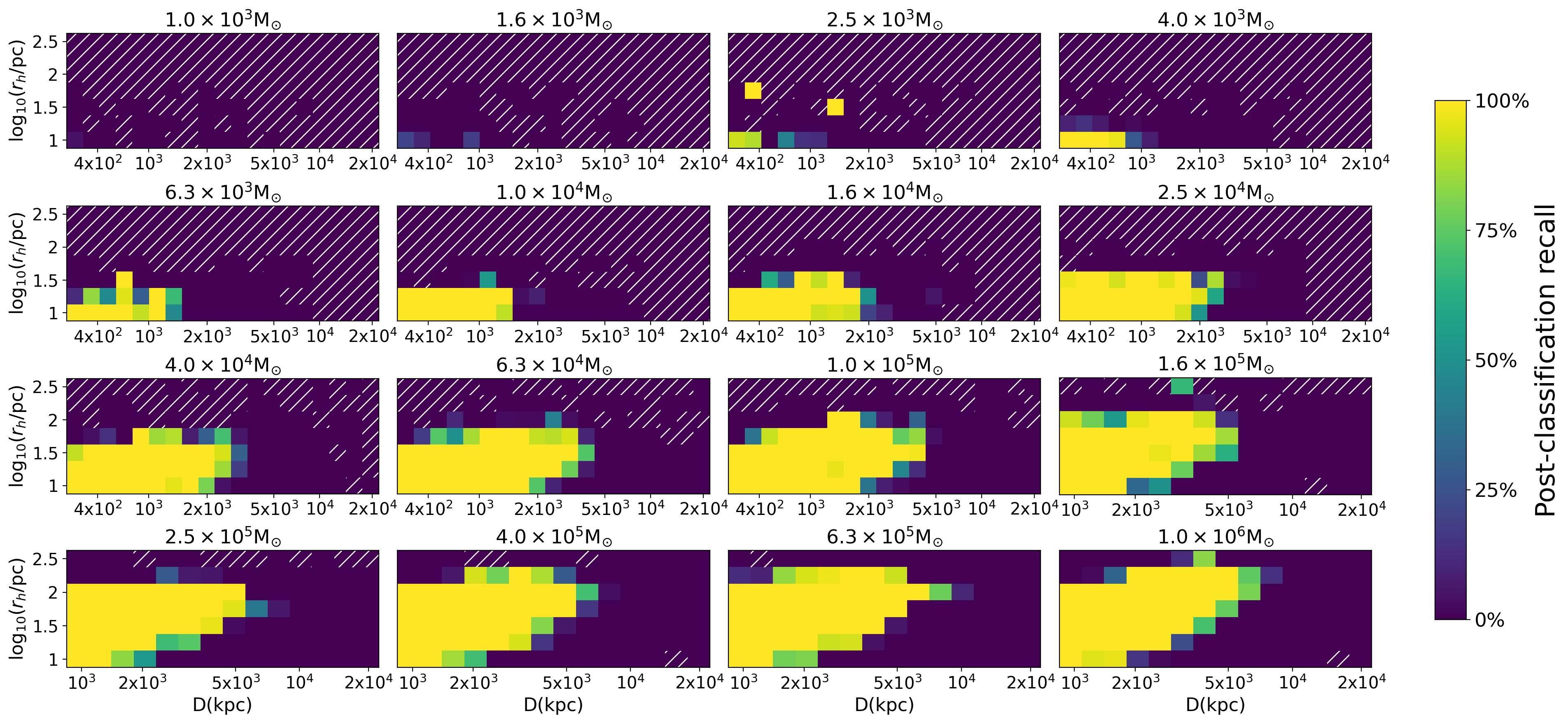}     
   \caption{Classification recall during the post-processing stage. The shaded regions indicate areas without test samples. }
   
   \label{fig:sexdetection}
\end{figure*}

\section{POST PROCESSING}
\label{sect:postprocess}

\begin{figure*}
   \centering
   \includegraphics[width=17.5cm, angle=0]{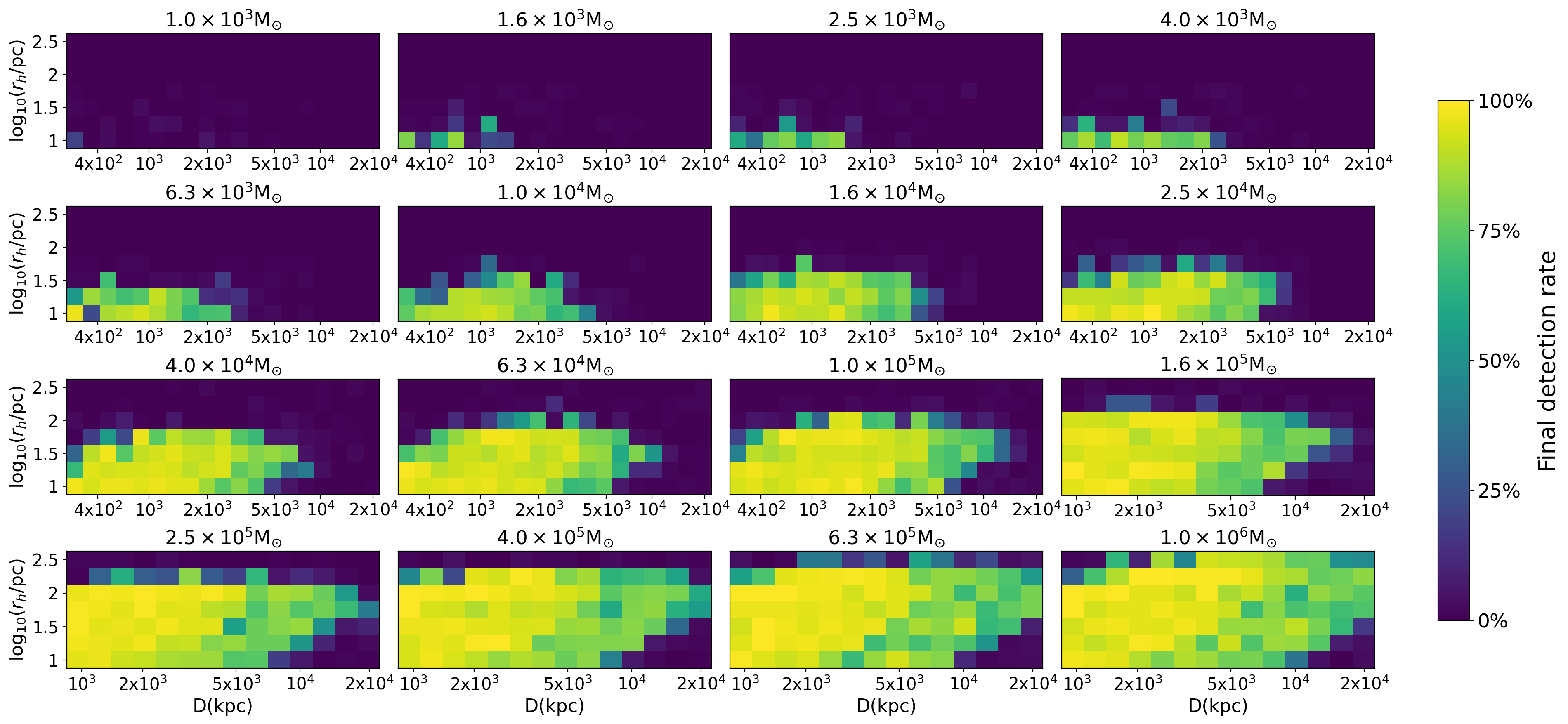}     
   \caption{Combined detection rate for LV dwarf galaxies, obtained by combining the ViT prediction results with the post-processing step (Figure~\ref{fig:vitp_detectivity}) and multiplying them with the detection rates from Step 1 (Figure~\ref{fig:step1_detectivity}). }
  
   \label{fig:final_detectivity}
\end{figure*}

\begin{figure*}
   \centering
   \includegraphics[width=15.5cm, angle=0]{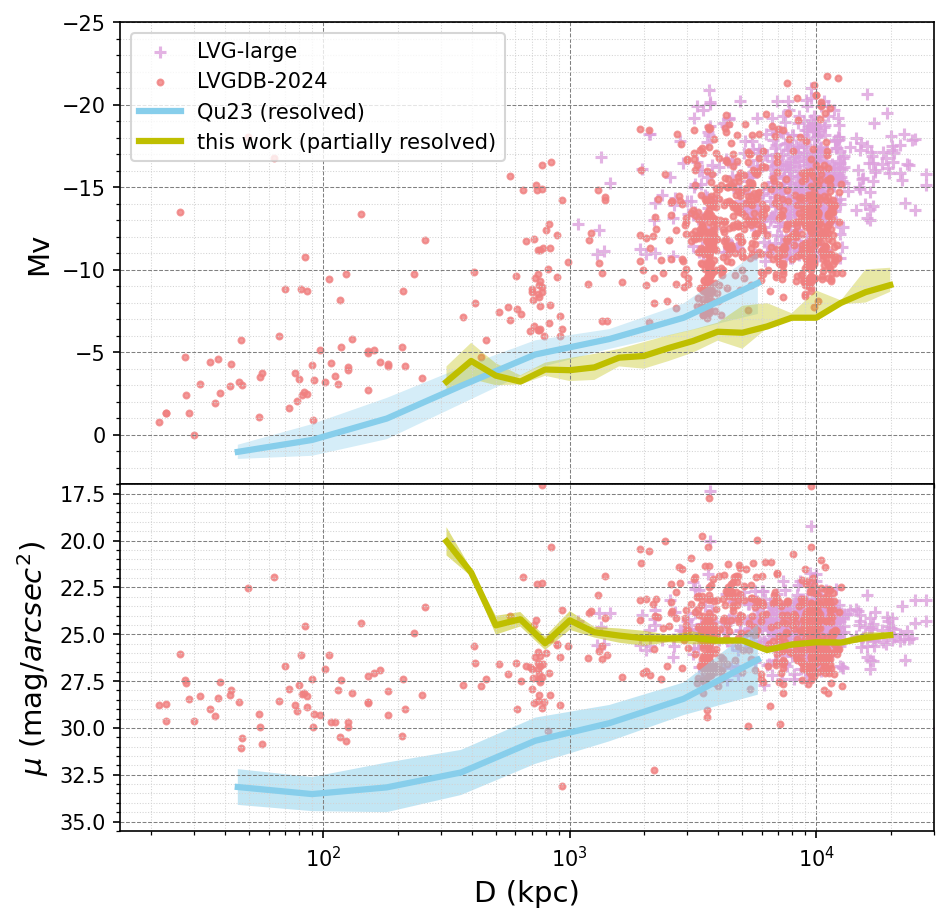}
   \caption{Comparison between the CSST dwarf galaxy detection limits derived in this work and the distribution of observed nearby galaxies.Top panel: Absolute magnitude as a function of distance. Bottom panel: Surface brightness versus distance.Purple plus signs and red circles indicate known Local Volume galaxies \citep{lvg2013,Pace2024}.The green line shows the detection limits from this study, with the solid line marking the 50\% completeness level and the shaded region spanning the 25\% to 75\% completeness range. The blue line indicates the detection limit from \citetalias{qu2023} for comparison.}
   \label{fig:cpfinal_detectivity}
\end{figure*}

In constructing the ViT model, we noticed that some LV dwarf galaxies are mis-classified, as shown in panel (a) in Figure~\ref{fig:postprocess} (more examples are shown in Figure~\ref{fig:false_dg}). A common feature observed in some of these galaxies is the presence of resolved member stars and a central concentration of stellar components. These features allow us to distinguish them from background galaxies (as shown in the second row of Figure~\ref{fig:show_candidates}), which generally lack such central overdensities. Motivated by this morphological distinction, we introduce a post-processing step to enhance the classification accuracy of LV dwarf galaxies.

In the post-processing, we re-examine all samples initially classified as negative by the ViT model, evaluating their central overdensities based on both the number of objects and their spatial concentration in the central region. This additional step allows us to identify potential LV dwarf galaxy candidates that are missed by the ViT classifier. As a result, the recovery rate of true LV dwarf galaxies can be improved without significantly increasing the FPR.

\subsection{Central overdensity evaluation}
\label{sect:method_postprocess}

The post-processing is performed on unscaled cutout images, with dimensions corresponding to the full region displayed in panel (h) of Figure~\ref{fig:detection}. For each sample, we use Source Extractor to identify sources in both the $g$- and $i$-band images, and retain only the matched detections, hereafter referred to as "SE-detections". We then assess potential source overdensities at the center of each cut-out.

The detailed procedure is illustrated in Figure~\ref{fig:postprocess}. Panel (a) displays the spatial distribution of SE-detections in the cutout image, with red circles indicating "SE-detections" identified in both the $g$- and $i$-band images. The cutout region is divided into 1"/pixel bins, and the number of SE-detections in each bin is used to construct the object density map (panel b). To enhance potential overdensity signals, this map is convolved with a Gaussian kernel of width $\sigma_1$ (panel c), while a broader Gaussian kernel ($\sigma_2$) is applied to generate a background reference (panel d). Subtracting the background from the smoothed density map yields the residual object density map (panel e). The S/N map is then obtained by normalizing the residual map with the standard deviation of the background, estimated from the outer two-thirds of the region (panel f). Finally, we measure two key quantities within the central third of each cut-out: the central object count (from Panel b) and the center S/N (from Panel f). These metrics serve as indicators of central source overdensities and are used in the subsequent classification refinement.

In Figure~\ref{fig:sn_postprocess}, we show the distributions of "central object count" and "central S/N" for all samples initially classified as negative by the ViT model, including both true positives (actual LV dwarf galaxy) and true negatives. The vertical axis indicates the percentage of samples. The results demonstrate that LV dwarf galaxy samples exhibit significantly higher central overdensities than true negatives. Based on this distinction, we apply the following selection criteria: samples with "central object count" > 7 and "central S/N" > 6 are reclassified as LV dwarf galaxies.

After applying the post-processing step, approximately 45\% of LV dwarf galaxies initially misclassified as negative by the ViT model are correctly reclassified as new positive. Meanwhile, the contamination rate (the proportion of true negatives incorrectly reclassified as LV dwarf galaxies) increases by only 0.02\%.

Figure~\ref{fig:sexdetection} shows the post-process detection rate distribution for LV dwarf galaxies. A clear trend emerges: brighter, nearer, and more spatially extended systems are more likely to be recovered. This is attributed to their higher number of resolvable member stars and denser spatial profiles, making them more distinguishable. These findings confirm that the post-process step significantly improves LV dwarf galaxy classification accuracy with minimal false positives. This serves as a complementary approach to the ViT-based classifier, which would optimize the detectability across different distances.
By incorporating the post-process results into the ViT classification output, the success rate of LV dwarf galaxy identification is further improved beyond what is shown in Figure~\ref{fig:vitdetection}. This enhancement is illustrated in Figure~\ref{fig:vitp_detectivity}.

\subsection{Overall detection rate}
\label{sect:final_result}

As previously outlined, our pipeline consists of three main steps: pre-processing, ViT classifer, and post-processing. By combining the detection and classification completeness achieved at each of these stages, we derive the overall detection efficiency of our LV Dwarf Galaxy Detection Pipeline, as shown in Figure~\ref{fig:final_detectivity}. This figure provides a comprehensive assessment of this work in identifying and classifying dwarf galaxies within the local volume.

Figure~\ref{fig:cpfinal_detectivity} compares our detection results with those of currently known nearby dwarf galaxies.
To represent the observational sample, we incorporate two Local Volume galaxy catalogs: LVG-large \citep{lvg2013} and LVGDB-2024 \citep{Pace2024}. The LVGDB-2024 catalog (marked by red circles in Figure~\ref{fig:cpfinal_detectivity}) serves as our primary reference, as it includes all known dwarf galaxies within 3 Mpc and reliable photometric measurements. The LVG-large catalog (indicated by pink plus signs) is used as a complementary dataset. Due to its broader coverage, we only include galaxies beyond 1 Mpc that are not already present in LVGDB-2024. Since LVG-large does not provide direct values for $M_{V}$ or surface brightness, we estimate them based on $a_{26}$ (the semi-major axis at the 26 mag/arcsec$^{2}$ isophote)) and $m_{26}$ (the integrated magnitude within that isophote), which may introduce small deviations from true values.

In our previous study (\citetalias{qu2023}), we employed a classic matched-filter technique to evaluate CSST's detection capabilities for Local Group dwarf galaxies, using only the simulated stellar catalogs. In the present work, we build upon that analysis by comparing the detection limits derived from both methods within overlapping parameter spaces. The green and blue lines in Figure~\ref{fig:cpfinal_detectivity} represent the detection limits obtained in this work and in \citetalias{qu2023}, respectively. Compared to existing observations, our detection limits reach fainter magnitudes at fixed distances within 20 Mpc, demonstrating the advantage of the image-based method proposed in this work for detecting smaller and more distant dwarf galaxies. However, beyond 3 Mpc, the surface brightness of known dwarf galaxies typically falls below our detection threshold of 25 mag/arcsec$^{2}$. These systems are primarily discovered through small-area, deep-exposure surveys targeting satellite populations, often around specific host galaxies. Such focused observations allow for extended exposure times and thus achieve deeper magnitude limits than CSST. 
In contrast, the CSST's main strength lies in its substantially larger sky coverage, enabling a broader and more uniform search for faint dwarf systems.

Compared to this work, \citetalias{qu2023} achieves higher recovery rates for nearby systems (owing to their larger angular sizes and more easily resolved stars) but it is less effective for distant or unresolved systems. Although it reaches a fainter surface brightness limit overall,  the method proposed in this work is more sensitive in terms of $M_V$. Notably, beyond 500 kpc, our detection limit in $M_V$ surpasses that of \citetalias{qu2023} by more than one magnitude. For galaxies at distances exceeding 5 Mpc, the absence of resolved bright stars further limits the applicability of the \citetalias{qu2023} algorithm. In contrast, our current approach remains effective in this regime, achieving higher detection rates and showing better suitability for detecting unresolved systems at greater distances.

We note the existence of two dwarf galaxies that lie well below our detection threshold: NGC55-dw1, a satellite of NGC 0055 \citep{McNanna2024} ($M_V = -8$, distance = 2.2 Mpc, $\mu = 32.25$ mag/arcsec$^{2}$), and Triangulum IV, a satellite of M33 \citep{Ogami2024} ($M_V = -6.39$, distance = 933 kpc, $\mu = 33.09$ mag/arcsec$^{2}$). Triangulum IV was discovered using HSC data, with a depth comparable to CSST. \cite{Ogami2024} reported a surface brightness of $\mu = 29.72$ mag/arcsec$^{2}$, which is above the CSST detection threshold; however, \cite{Pace2024} reports a deeper value of $\mu = 33.09$ mag/arcsec$^{2}$. The other system, NGC55-dw1, was identified in DES Year 6 data, using an improved matched-filter algorithm that focused on the 300 kpc to 2 Mpc range. That study introduced refined color filtering and assessed local stellar overdensities across multiple scales, achieving a significantly deeper detection threshold. The detection strategy employed in that work (particularly the methodological refinements) offers useful insights for future CSST-based searches. 

\section{Summary}

\label{sec:sum}

The search and identification of dwarf galaxies in the Local Volume are crucial for constructing the satellite galaxy luminosity function in nearby systems. To carry out a comprehensive and effective search, the depth and quality of observational data, as well as the efficiency of the detection algorithm, are all essential.
The upcoming CSST sky survey provides a new opportunity for the comprehensive search for nearby dwarf galaxies. In this study, we systematically evaluate the detection capabilities of CSST for dwarf galaxies within the Local Volume.

Using the CSST Image Simulator, we generated multi-band synthetic images based on the primary survey parameters of CSST, along with a set of mock images representing LV dwarf galaxies spanning a wide range of distances, magnitudes, and structural properties. We developed a three-step detection pipeline consisting of extended source detection from images, and classification using a ViT model to identify LV dwarf galaxies. For dwarf galaxy systems misclassified by the ViT model, the developed post-processing steps can recover around half of them. Within this framework, we quantified the detection and classification completeness for dwarf galaxies in the CSST Wide Survey.

The classification component in our pipeline employs the "vit-base-patch16-224-in21k" model, which has demonstrated strong performance in identifying nearby dwarf galaxies from simulated CSST imaging data in this work, achieving a true positive rate (TPR) exceeding 85\% at a fixed false positive rate (FPR) of 0.1\%. To further improve completeness, a post-process step is introduced to re-examine initially negatively misclassified dwarf systems, enabling the recovery of originally missed LV dwarf galaxy candidates. This step increases the overall TPR to approximately 92\%, without a significant increase in FPR.

In comparison with the method proposed in \citetalias{qu2023}, we find that the two algorithms exhibit complementary detection capabilities across the parameter space defined by galaxy distance, half-light radius, and absolute magnitude. While \citetalias{qu2023} performs more effectively for nearby galaxies with larger angular sizes as well as more easily resolvabled stars, the approach developed in this work is better suited for detecting distant or incompletely resolved systems. These differences underscore the value of combining both techniques to achieve a more complete census of dwarf galaxies in the Local Volume. Based on the detection efficiency derived from our pipeline, we find that a 50\% detection rate corresponds to a limiting absolute magnitude of $M_V \lesssim -7$ within 10 Mpc, and a surface brightness threshold of $\mu \gtrsim 25$ mag/arcsec$^{2}$ beyond 1 Mpc. These results indicate that our method is well suited for enabling a systematic search for ultra-diffuse galaxies (UDGs) across the Local Volume.

We emphasize that the detection algorithm proposed in this study is fully image-based and does not rely on any higher-level data products. The only preprocessing steps required are reference image calibration and image stacking. This allows the method to be applied directly to single-exposure pointings. This enables real-time detection and candidate identification during the early stages of the CSST survey. In contrast, the approach presented in \citetalias{qu2023} depends on pre-processed stellar catalogs, underscoring the flexibility and operational independence of our pipeline.

In practice, real observational data present additional complexities not captured in simulations, include higher densities of large background galaxies, increased noise from instrumental and atmospheric effects, and diffuse light contamination from nearby host galaxies. These factors may affect the completeness and reliability of source detection and classification. To address this, our pipeline incorporates several tunable parameters such as the S/N and size thresholds for extended source detection. The strategies for constructing the training dataset can be further optimized once real CSST data become available.

Although our method is designed to be fully independent of external data products, integrating it with supplementary information such as star-galaxy classification, photometric redshifts, or multi-band photometry may substantially enhance detection performance. This hybrid approach could be especially valuable for identifying marginal or ultra-diffuse systems, and represents a promising direction for maximizing the scientific return of CSST in the study of low surface brightness galaxies.



\section*{Acknowledgements}
We acknowledge the science research grants from the China Manned Space Project with NO. CMS-CSST-2025-A11 and CMS-CSST-2025-A20, the cosmology simulation database (CSD) in the National Basic Science Data Center (NBSDC) and its funding NBSDC-DB-10 (No.2020000088). We utilized the high-performance computing cluster from the Purple Mountain Observatory of the Chinese Academy of Sciences, which is equipped with GPU acceleration cards made in China.

X.K. acknowledge the support from the National Key Research and Development Program of China (No. 2022YFA1602903), the China Manned Space project with NO.CMS-CSST-2025-A10

\bibliographystyle{mnras}
\bibliography{ref.bib} 
\appendix
\section{ADDITIONAL FIGURES}
\begin{figure*}
   \centering
   \includegraphics[width=15.5cm, angle=0]{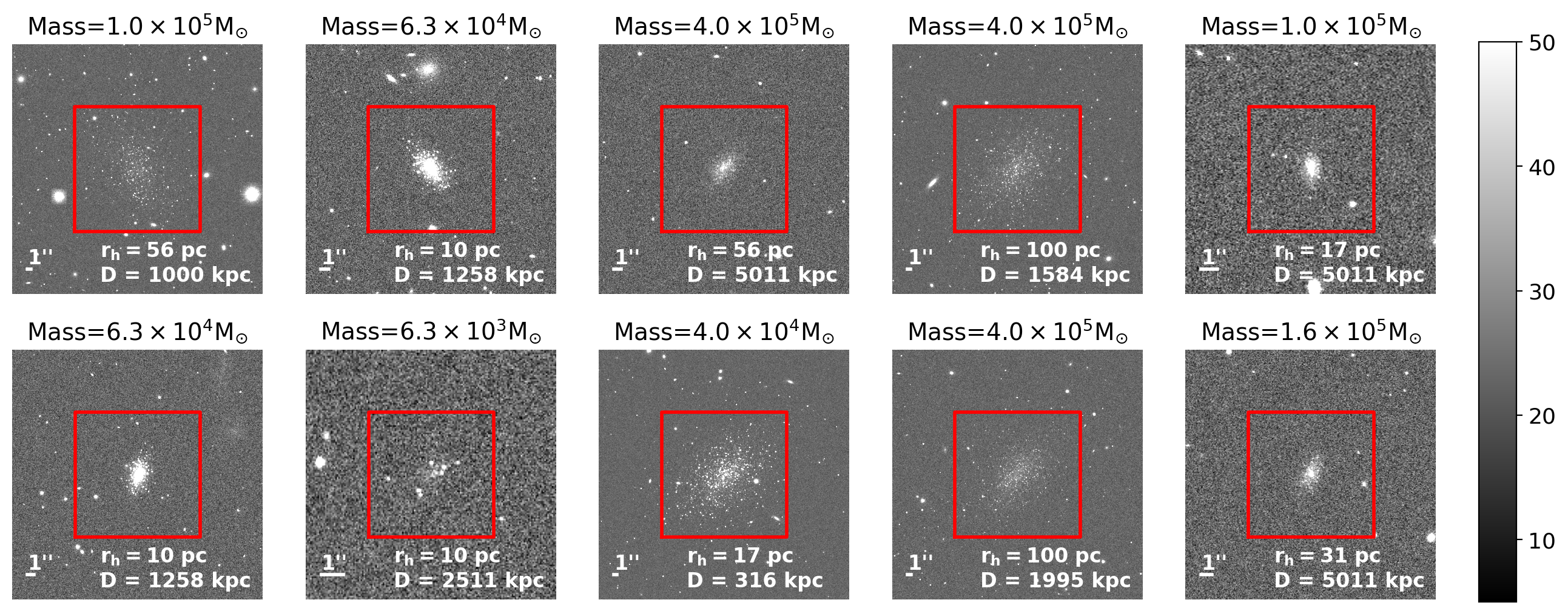}     
   \caption{Unscaled cutouts (in $g$-band) of LV dwarf galaxies misclassified by the ViT model.
   }
   \label{fig:false_dg}
\end{figure*}

\begin{figure*}
   \centering
   \includegraphics[width=16.5cm, angle=0]{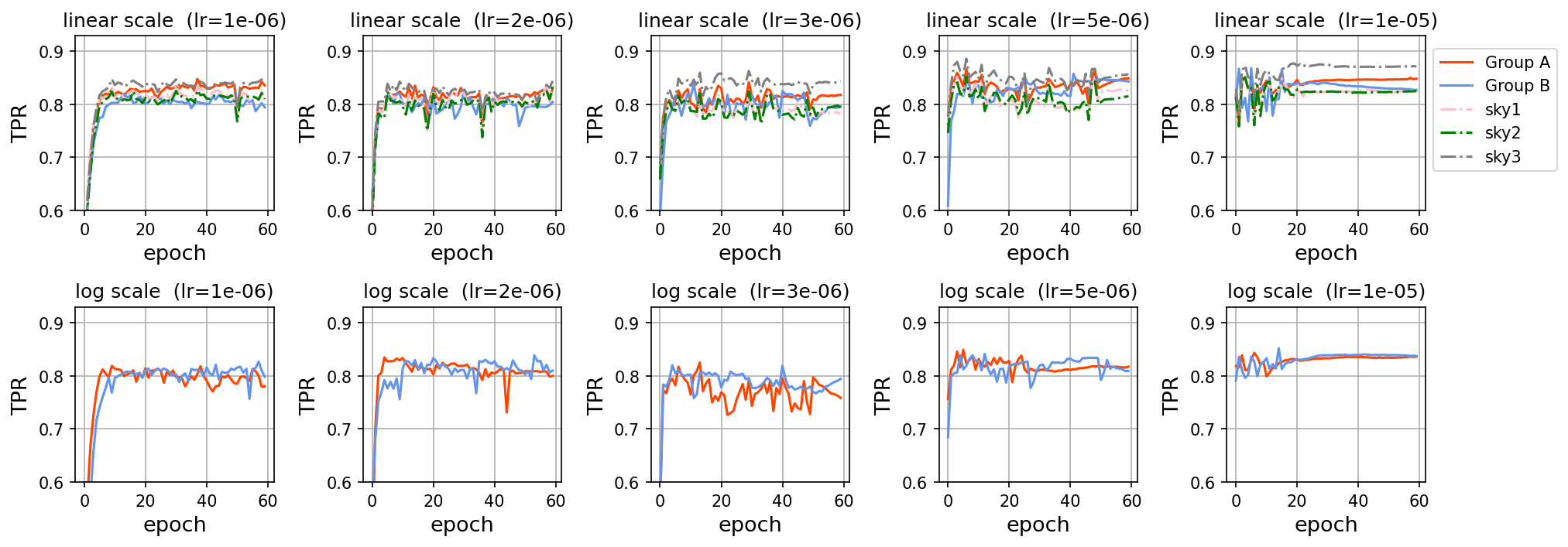}     
   
   \caption{The variation of TPR at FPR=0.001 across different testing sets with training epochs under various learning rates. The first row presents the results for cut images processed with linear scaling, while the second row shows the results for cut images processed with logarithmic scaling. Each column corresponds to a different learning rate.
   }
   \label{fig:tpr_vit}
\end{figure*}

\begin{figure*}
   \centering
   \includegraphics[width=17.5cm, angle=0]{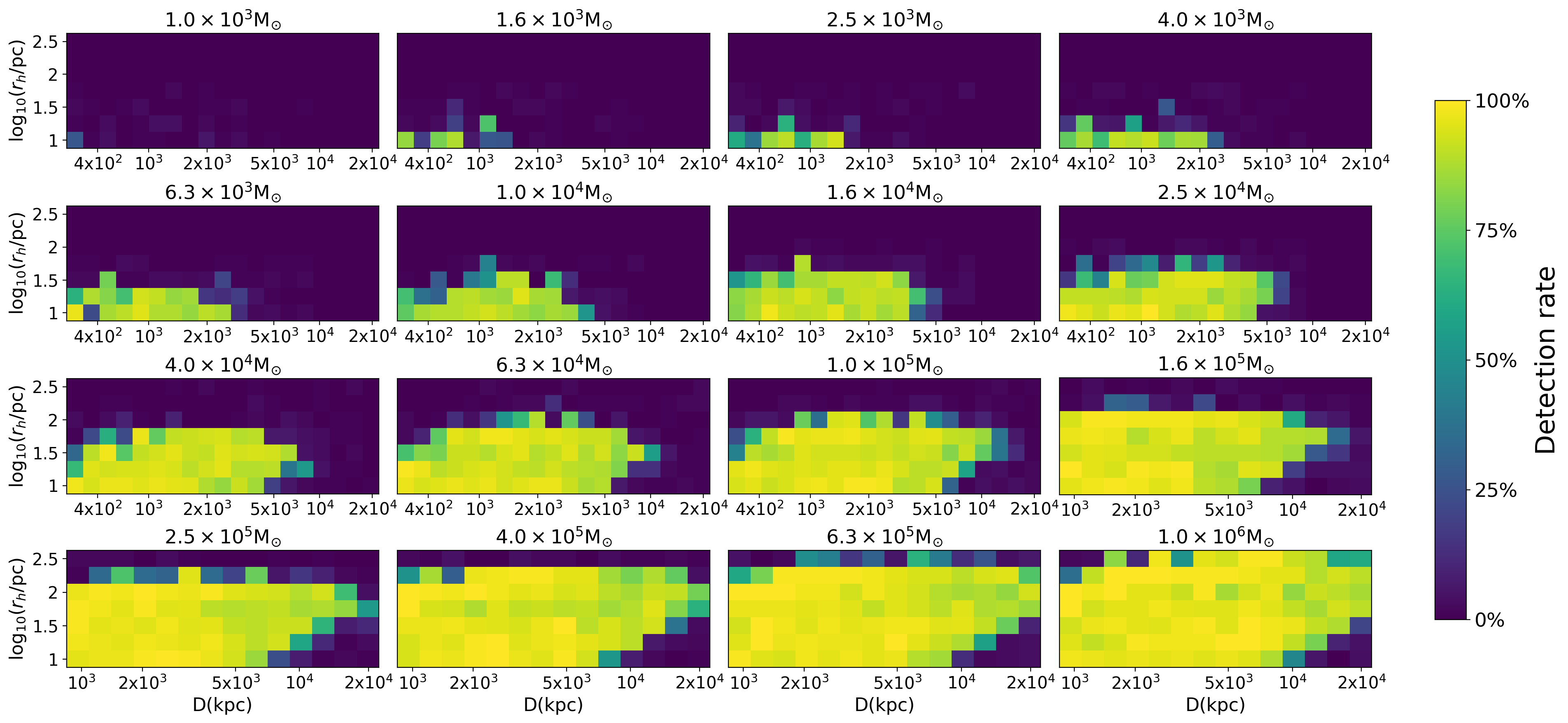}     
   \caption{Detection rates of LV dwarf galaxies in the extended source detection step within pre-process.}
  
   \label{fig:step1_detectivity}
\end{figure*}

\begin{figure*}
   \centering
   \includegraphics[width=17.5cm, angle=0]{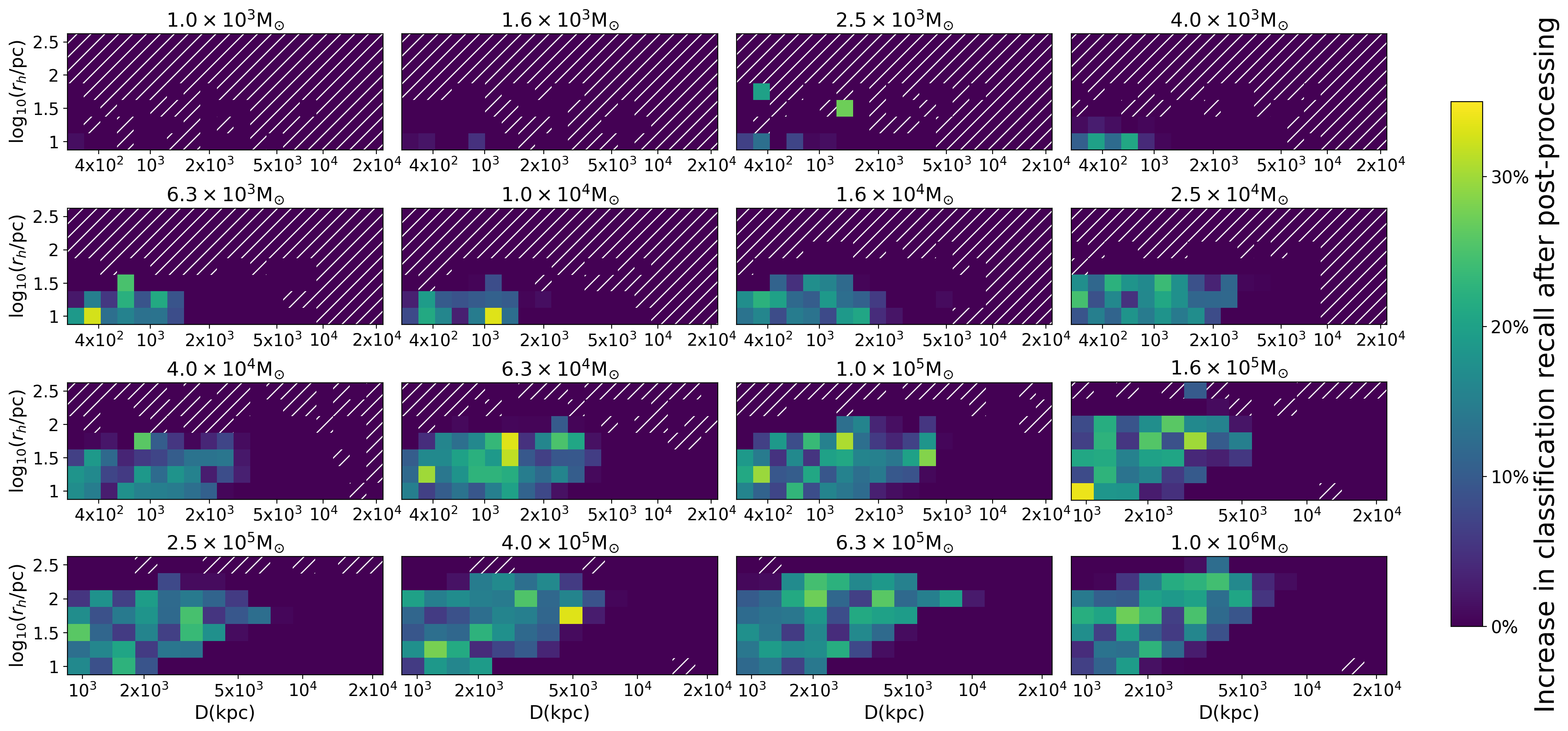}     
   \caption{Enhancement of the classification recall through post-process}
   \label{fig:vitp_detectivity}
\end{figure*}


\bsp	
\label{lastpage}
\end{document}